\documentclass[aps,superscriptaddress,notitlepage,prx,reprint]{revtex4-2}
\usepackage{epsfig,color}
\usepackage{graphicx}
\usepackage{dcolumn}
\usepackage{bm}
\usepackage{amsmath,amsfonts,amssymb,mathrsfs}
\usepackage{pstricks}
\usepackage{amsxtra}
\usepackage{cases}
\usepackage{amsthm}
\usepackage{braket}
\usepackage{natbib}
\usepackage{physics}
\usepackage{hyperref,color}
\usepackage{comment}
\definecolor{darkblue}{rgb}{0.0,0.0,0.7}
\hypersetup{colorlinks,breaklinks,linkcolor=darkblue,urlcolor=darkblue,anchorcolor=darkblue,citecolor=darkblue}

\begin{document}
\title{Analog Quantum Phase Estimation with Single-Mode Readout}

\author{Wei-Chen Lin}
\affiliation{Department of Physics and Center for Theoretical Physics, National Taiwan University, Taipei 106319, Taiwan}

\author{Chiao-Hsuan Wang}
\email{chiaowang@phys.ntu.edu.tw}
\affiliation{Department of Physics and Center for Theoretical Physics, National Taiwan University, Taipei 106319, Taiwan}
\affiliation{Center for Quantum Science and Engineering, National Taiwan University, Taipei 106319, Taiwan}
\affiliation{Physics Division, National Center for Theoretical Sciences, Taipei 106319, Taiwan}

\begin{abstract}
Eigenvalue estimation is a central problem for demonstrating quantum advantage, yet its implementation on digital quantum computers remains limited by circuit depth and operational overhead. We present an analog quantum phase estimation (aQPE) protocol that extracts the eigenenergies of a target Hamiltonian via continuous time evolution and single-mode cavity measurement. By encoding eigenvalue information as conditional cavity phase-space rotations, the scheme avoids deep quantum circuits and entangling gates, while enabling readout through established cavity tomography techniques. We further illustrate the feasibility of this approach by engineering a Hamiltonian that implements aQPE of the XY model, whose ground-state energy problem is QMA-complete, within a physical architecture compatible with existing circuit quantum electrodynamics technology. Our results provide a resource-efficient and scalable framework for implementing quantum phase estimation in near-term quantum platforms.
\end{abstract}
\maketitle

\section{Introduction}\label{sec:intro}
Quantum computers promise to solve classically intractable problems, with eigenvalue estimation as a prominent avenue for demonstrating quantum advantage. Determining the eigenvalues of quantum operators presents a central challenge across multiple disciplines, including extracting molecular energies in quantum chemistry~\cite{Lanyon2010,Cao2019,McArdle2020}, unveiling emergent features in materials science~\cite{Wecker2015,Babbush2018,Bauer2020}, probing many-body phase transitions~\cite{Sachdev2000,Roushan2017}, and enabling quantum machine learning protocols~\cite{Biamonte2017}. Due to the exponential growth of Hilbert space with system size, classical algorithms face fundamental limitations in solving eigenvalue problems, as representing large-scale quantum states quickly becomes intractable~\cite{Coleman1963,Liu2007}. 
Approximate numerical methods, such as density functional theory~\cite{Schuch2009,Jones2015} and quantum Monte Carlo~\cite{Foulkes2001,Luchow2011,Carlson2015}, are often unreliable in the presence of strong correlations or entanglement.  In light of classical limitations, quantum computers have emerged as a natural platform for eigenvalue estimation.
\vspace{1.7em}

Using digital quantum computers, eigenvalue problems can be efficiently solved with the quantum phase estimation (QPE) algorithm, implemented through long quantum circuits~\cite{Kitaev1995,Cleve1998,Abrams1999}.  QPE has enabled the extraction of eigenenergies in many-body quantum systems, including quantum chemistry simulations~\cite{Aspuru2005}. By estimating the phase accumulated under a unitary evolution $\hat{U} = e^{- i \hat{H}_E t/\hbar}$, the algorithm effectively determines the eigenenergies of the underlying Hamiltonian operator $\hat{H}_E$.  Beyond this application, QPE also serves as a key subroutine in several algorithms, such as enabling exponential speedup in Shor’s factorization~\cite{Shor1999}.  However, the implementation of QPE algorithm typically requires deep circuits and Trotterization~\cite{Trotter1959} of the unitary evolution, posing demanding requirements on current quantum hardware.

To overcome the hardware limitations of standard QPE, previous approaches have focused on algorithmic improvements, variational methods, and analog simulations. Variants of QPE have been developed to reduce circuit depth and enhance robustness~\cite{Griffiths1996,Wiebe2016,Russo2021,Ni2023}, while still relying on complex entangling operations such as controlled-unitaries or quantum Fourier transforms. The variational quantum eigensolver (VQE)~\cite{Peruzzo2014,McClean2016,Tilly2022}, on the other hand, circumvents deep circuits and complex quantum operations by classical optimization over parameterized ansatz states, at the expense of scalability and measurement overhead. In contrast, analog quantum simulators access spectral features via continuous evolution~\cite{Senko2014,Roushan2017}, while facing challenges in measurement and verification~\cite{Hauke2012}.

Here, we present a practical analog quantum phase estimation (aQPE) protocol that extracts the eigenenergies of a target Hamiltonian through continuous evolution and single-mode measurement. Inspired by the success of dispersive readout in superconducting circuit quantum electrodynamics (QED)~\cite{Wallraff2005,Blais2007}, our scheme encodes the eigenenergies as conditional phase-space rotations of a single cavity state. This approach not only eliminates the need for deep circuits and entangling operations in digital quantum computers but also enables readout mechanisms that remain challenging for analog quantum simulators. We introduce a Hamiltonian engineering method to implement aQPE for the XY model, whose ground-state energy problem is known to be QMA-complete~\cite{Kempe2006,Cubitt2016,Piddock2017}, and show that this protocol can be directly realized using existing circuit QED technology. This architecture offers a resource-efficient pathway toward scalable quantum phase estimation in near-term devices.

\section{Analog Quantum Phase Estimation}
\begin{figure}[htbp]
\begin{center}
\includegraphics[width=\linewidth]{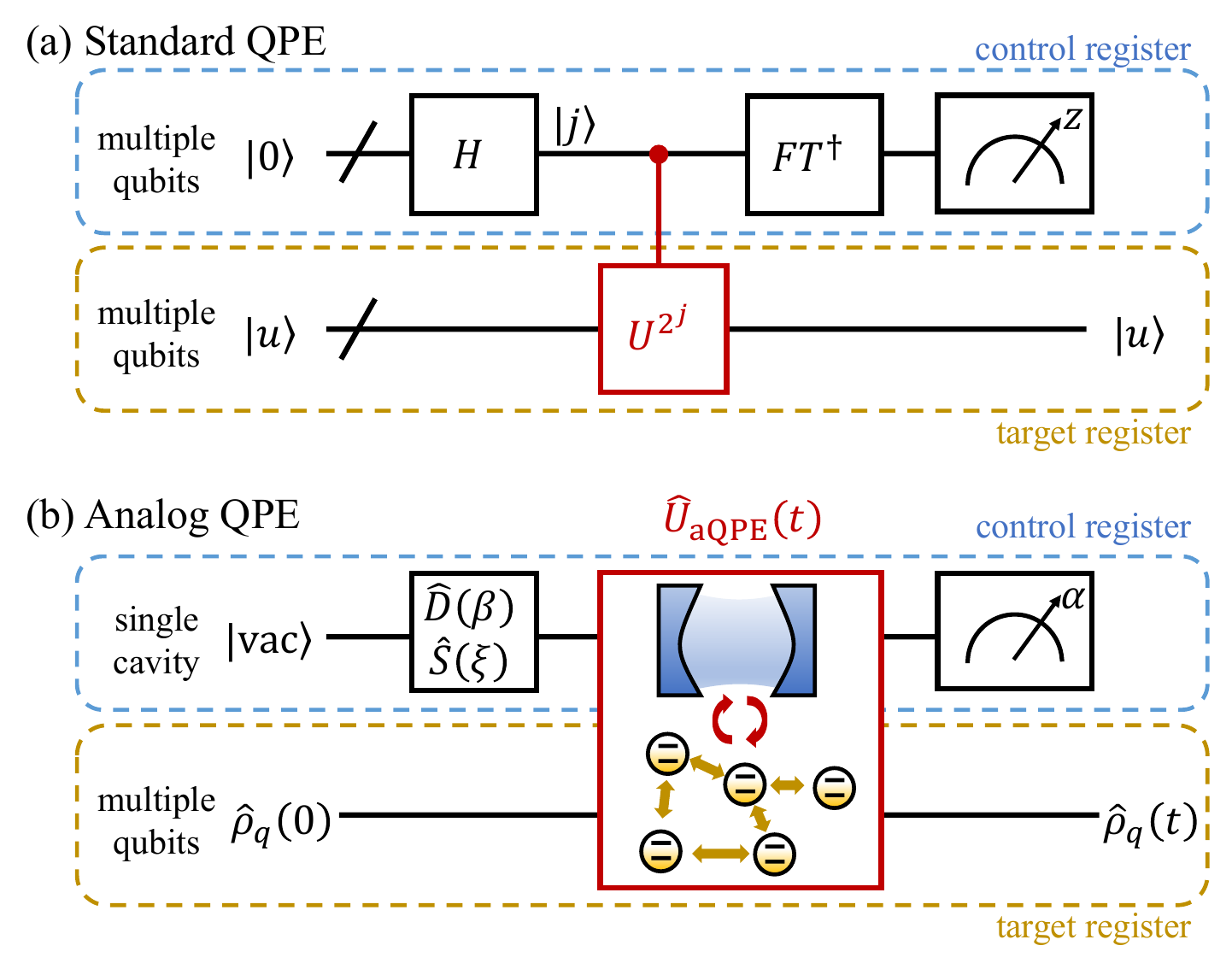}
\caption{Correspondence between standard and analog quantum phase estimation. (a) Circuit diagram of the standard QPE algorithm~\cite{Nielsen2010}.  A control register is initialized in a uniform superposition state by Hadamard gates and a multi-qubit target register is prepared in an eigenstate $\ket{u}$ of a unitary operator $\hat{U}$. A sequence of controlled-$U^{2^j}$ operations imprints the eigenphase of the target register onto the control register. The control register is then processed by an inverse QFT and measured in the computational basis to extract the phase information.  (b) Schematic of the analog QPE protocol. A cavity (control register) is initialized in a phase-squeezed coherent state, and multiple qubits (target register) are prepared in an arbitrary state $\hat{\rho}_q(0)$.  Continuous evolution $\hat{U}_{\rm aQPE}(t)$ imprints the spectral information of the multi-qubit state onto the cavity.  The cavity is then measured in phase space to extract the eigenenergies and corresponding populations of the initial target register state, with respect to a given target Hamiltonian.}
\label{fig:Schematics}
\end{center}
\end{figure}

We propose an analog quantum phase estimation (aQPE) protocol, in which the phase-space evolution of a cavity mode encodes the spectral information of a multi-qubit Hamiltonian problem. To motivate the analog approach, we begin by reviewing the standard circuit-based quantum phase estimation (QPE) algorithm, as shown in Fig.~\ref{fig:Schematics}(a). In standard QPE, a multi-qubit control register is initialized in a uniform superposition state by Hadamard gates, and a multi-qubit target register is prepared in an eigenstate $\ket{u}$ of a unitary operator
$\hat{U}=e^{- i \hat{H}_E t/\hbar}$. A sequence of controlled-$U^{2^j}$ gates imprints the eigenphase of the target register onto the control register, followed by an inverse quantum Fourier transform (QFT) and measurements to extract the eigenphase and corresponding eigenenergy of $\hat{H}_E$.

In our analog protocol illustrated in Fig.~\ref{fig:Schematics}(b), the multi-qubit control register in standard QPE is replaced by a single-mode cavity, and its joint evolution with the multi-qubit target system serves as the analog counterpart to the controlled-unitary gates. For a given target Hamiltonian $\hat{H}_{\rm target}$ acting on the multi-qubit target register, we design a cavity-multiqubit Hamiltonian in dispersive form,
\begin{align}
    \hat{H}_{\rm{aQPE}} = \hat{a}^{\dagger} \hat{a} \otimes \hat{H}_{\rm target},
\label{eqn:HaPQE}
\end{align}
where $\hat{a}^{\dagger}$ and $\hat{a}$ are the creation and annihilation operators of the cavity mode. This Hamiltonian effectively induces a qubit-state-dependent frequency shift for the cavity, resulting in a unitary evolution after time~$t$:
\begin{align}
    \hat{U}_{\rm aQPE}(t) = \sum_n \ket{n}\bra{n} \otimes \left( \hat{U}_{\rm target}^n \right),
    \label{eqn:UaQPE}
\end{align}
where $\ket{n}$ denotes the $n$-photon Fock state, and $\hat{U}_{\rm target} = e^{-i \hat{H}_{\rm target} t / \hbar}$. Conditioned on the cavity photon number $n$, the designed Hamiltonian applies a unitary evolution $\hat{U}_{\rm target}^n$ on the target register, effectively raising the unitary to a photon-number-dependent power.  This evolution mirrors a sequence of controlled-$U^{2^j}$ in standard QPE, imprinting the spectral information of the qubits onto the cavity.

The spectral information encoded by the continuous-time evolution $\hat{U}_{\rm aQPE}(t)$ can be directly accessed through typical preparation and measurement techniques applied to the cavity state.  We consider an initial state of the total cavity-multiqubit system described by the total density operator $\hat{\rho}_0 = \hat{\rho}_{c}(0) \otimes \hat{\rho}_{q}(0)$. The target register is prepared in a general (possibly mixed) state $\hat{\rho}_{q}(0) = \sum_{i,j} p_{ij} \ket{e_i}\bra{e_j}$, where $\ket{e_j}$ are the eigenstates of $\hat{H}_{\mathrm{target}}$ with corresponding eigenenergies $E_j$. The cavity is initialized in a phase-squeezed coherent state $\hat{\rho}_{c}(0) = \ket{\beta, \xi} \bra{\beta, \xi}$, where $\ket{\beta, \xi} = \hat{D}(\beta)\hat{S}(\xi)\ket{\rm vac}$, and $\hat{D}(\beta)$ and $\hat{S}(\xi)$ are the displacement and squeezing operators, respectively, and $\ket{\rm vac}$ denotes the cavity vacuum~\cite{Walls2008}. The displacement operation naturally generates a superposition over Fock states, paralleling the Hadamard-initialized control register in standard QPE, while the optional phase-squeezing operation can reduce angular uncertainty in the cavity phase space.

After a time evolution under $\hat{H}_{\mathrm{aQPE}}$, the reduced density operator of the cavity state becomes~\cite{note1}
\begin{align}
&\hat{\rho}_c(t)= \mathrm{Tr}_q \left[\hat{\rho}(t)  \right]= \mathrm{Tr}_q \left[\hat{U}^{\phantom{\dagger}}_{\rm aQPE}(t)\hat{\rho}_0 \hat{U}_{\rm aQPE}^{\dagger}(t) \right]\notag\\
=&\sum_j p_{jj} \ket{\beta e^{-iE_j t/\hbar}, \xi e^{ -i2E_j t/\hbar}} \bra{\beta e^{-iE_j t/\hbar}, \xi e^{-i2E_j t/\hbar}},
\label{eqn:rhoct}
\end{align}
where each term corresponds to a squeezed state component rotated by an angle $\Delta \theta_j = -E_j t/\hbar$ in phase space, with weight $p_{jj} = \bra{e_j} \hat{\rho}_{q}(0) \ket{e_j}$ reflecting the initial population of the target register in $\ket{e_j}$. The designed aQPE evolution thus induces a conditional rotation of the cavity state in phase space, with rotation angles determined by the eigenenergies of $\hat{H}_{\rm target}$ and amplitudes set by the initial state populations. The resulting cavity distribution can be directly accessed through phase-space tomography~\cite{Smithey1993,Lutterbach1997,Bertet2002,Krisnanda2025}, which naturally serves as the counterpart to the readout step in standard QPE, where a complex inverse QFT is typically required.

\begin{figure}[htbp]
\begin{center}
\includegraphics[width=\linewidth]{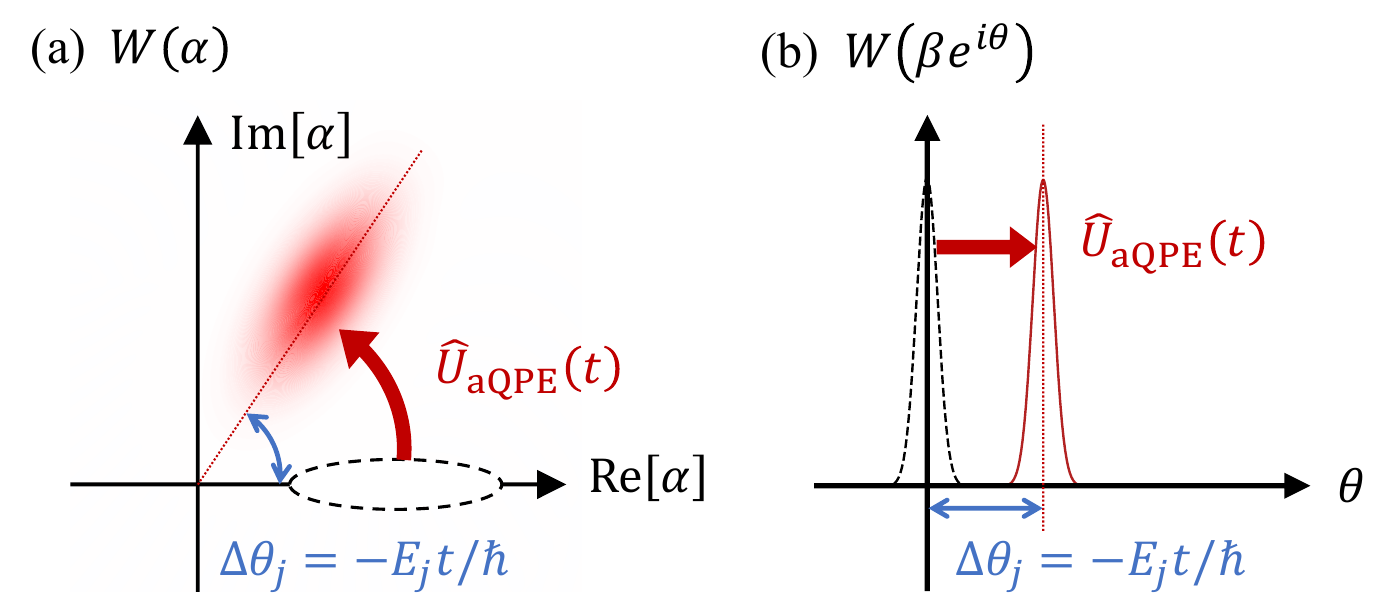}
\caption{Phase-space encoding of eigenenergies in the aQPE protocol. (a) Schematic evolution of the cavity Wigner function $W(\alpha)$ in phase space. The cavity is initialized at $t = 0$ in a phase-squeezed coherent state $\ket{\beta, \xi}$ with $\beta \in \mathbb{R}$ and $\xi < 0$, and is coupled to a multi-qubit register prepared in a single eigenstate $\ket{e_j}$ of the target Hamiltonian. Under the aQPE evolution,  $\hat{U}_{\rm aQPE}(t)$, the cavity distribution rotates by an angle $\Delta\theta_j = -E_j t/\hbar$, where $E_j$ is the eigenenergy associated with $\ket{e_j}$.  
(b) Schematic evolution of the angular profile of the Wigner function at fixed radius $\beta$, $W(\beta e^{i\theta})$, showing the corresponding energy-dependent rotation angle $\Delta\theta_j$. In both panels, the initial cavity distribution is indicated by a black dashed contour.}
\label{fig:Rotation}
\end{center}
\end{figure}

To illustrate the conditional rotation mechanism underlying aQPE, we show in Fig.~\ref{fig:Rotation} how eigenenergies are encoded as rotations in the cavity phase space. For simplicity, we consider a target register prepared in a single eigenstate $\ket{e_j}$ with eigenenergy $E_j$ of the target Hamiltonian $\hat{H}_{\rm target}$, which is coupled to a cavity initialized in a phase-squeezed coherent state $\ket{\beta, \xi}$. The cavity phase-space distribution can be visualized by its Wigner function $W(\alpha)$~\cite{Cahill1969}, illustrated schematically in Fig.~\ref{fig:Rotation}(a), with the initial state $\ket{\beta, \xi}$ represented by a black dashed contour centered at $\beta$ with reduced uncertainty along the angular direction (phase-squeezing).
Under the aQPE evolution $\hat{U}_{\rm aQPE}(t)$, the cavity distribution undergoes an angular rotation in the phase space by an angle $\Delta\theta_j = -E_j t/\hbar$, directly encoding the eigenenergy $E_j$. The eigenenergy $E_j$ can thus be extracted from the rotation angle by evaluating the Wigner function along a fixed radius, $W(\beta e^{i\theta})$, as shown in Fig.~\ref{fig:Rotation}(b).

Our aQPE protocol mirrors the structure of standard QPE, with the cavity initialization emulating the Hadamard-prepared superposition of the control register, the conditional phase-space rotation replacing controlled unitaries, and the phase-space tomography substituting for inverse QFT and projective readout. While the previous illustration focuses on a single eigenstate and a phase-squeezed coherent cavity state, the protocol applies more generally to arbitrary target states and a broader class of cavity preparations, with phase-squeezing being an optional operation for improving resolution. This analog scheme offers a resource-efficient and experimentally accessible approach to quantum phase estimation and can be implemented using available preparation and measurement techniques across diverse platforms, including cavity and circuit QED~\cite{Haroche1989,Raimond2001,Haroche2020,Blais2021,Ma2021}.

\section{XY model Spectral Estimation}\label{sec:XYModel}
We now investigate how the proposed aQPE protocol can be applied to extract the eigenenergies of the XY model through engineered cavity–multiqubit interactions.
We consider a target Hamiltonian given by a generalized XY model with symmetric, site-dependent couplings:
\begin{align}
\hat{H}_{\rm target}^{\rm XY} 
&= \sum_{\langle k,k'\rangle} 
\frac{\hbar\eta_{kk'}}{2} \left( 
\hat{\sigma}_k^{x} \hat{\sigma}_{k'}^{x} + 
\hat{\sigma}_k^{y} \hat{\sigma}_{k'}^{y} 
\right) \notag\\
&= \sum_{\langle k,k'\rangle} 
\hbar\eta_{kk'} \left(
\hat{\sigma}_k^{+} \hat{\sigma}_{k'}^{-} +
\hat{\sigma}_k^{-} \hat{\sigma}_{k'}^{+}
\right),
\label{eqn:HXY}
\end{align}
where $\hat{\sigma}_k^{x/y/z}$ denote the Pauli operators acting on the $k$-th qubit, and $\hat{\sigma}_k^{\pm} = (\hat{\sigma}_k^{x} \pm i \hat{\sigma}_k^{y})/2$ are the corresponding qubit raising and lowering operators. The notation $\langle k,k'\rangle$ denotes a pair of nearest-neighbor target qubits coupled at a strength $\eta_{kk'}$.  
This model features nearest-neighbor exchange interactions between quantum spins and serves as a paradigmatic example of quantum many-body systems~\cite{Lieb1961,Barouch1970,Kosterlitz1973}. Determining its ground-state energy is known to be QMA-complete~\cite{Kempe2006,Cubitt2016,Piddock2017}, highlighting its computational complexity.

\begin{figure}[htbp]
\begin{center}
\includegraphics[width=\linewidth]{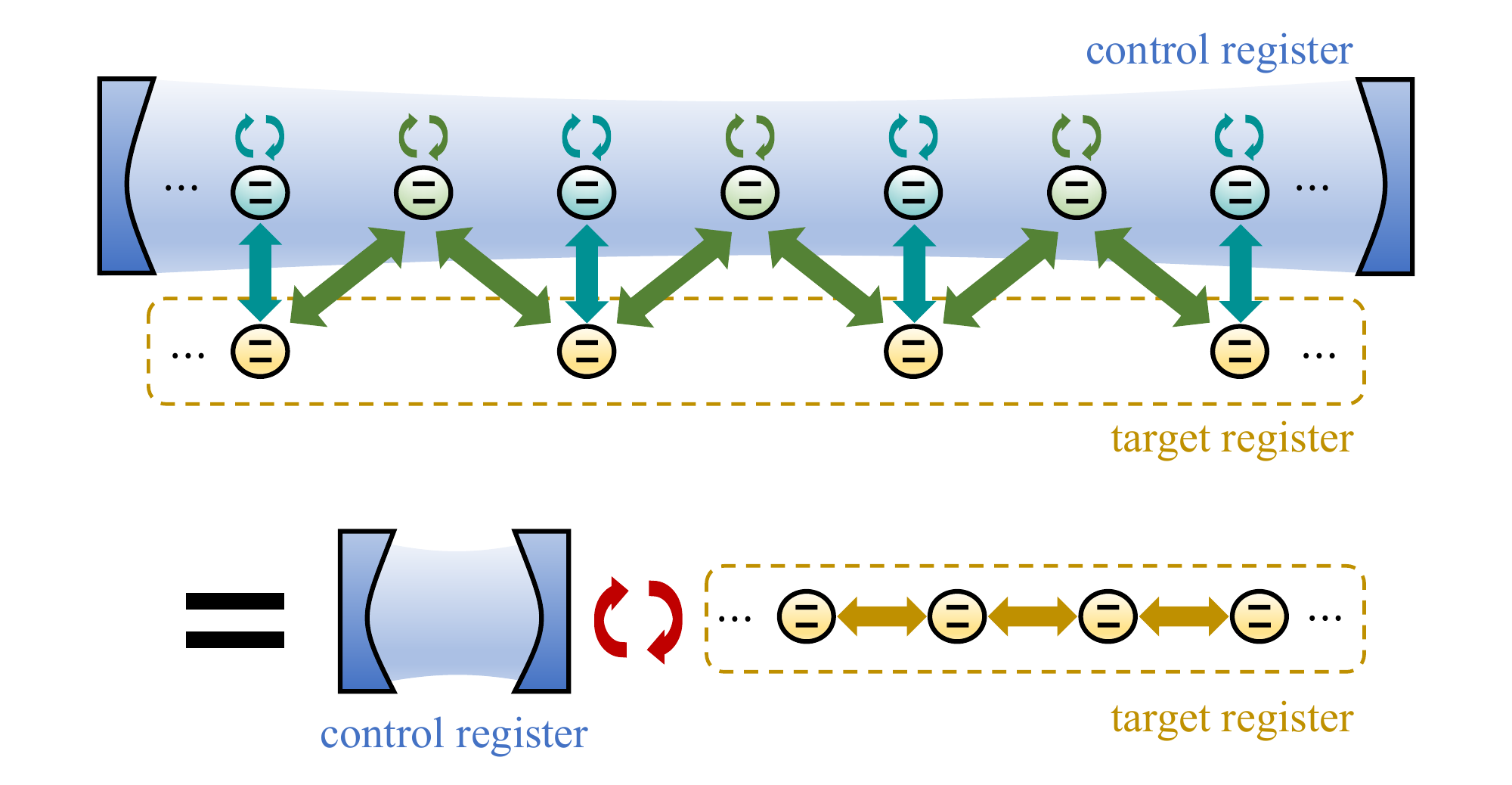}
\caption{Cavity–multiqubit architecture for implementing analog quantum phase estimation of the XY model.  The upper panel shows the physical layout, where yellow qubits within the dashed box comprise the target register. Dispersive coupling between the target register and the control cavity register is mediated by auxiliary coupler qubits: a local coupler qubit (teal) is coupled to a single target qubit, while a cross coupler qubit (green) is coupled to a fixed pair of target qubits and thereby determines the connectivity of the resulting XY model. The lower panel represents the engineered aQPE Hamiltonian structure in the dispersive form, 
$\hat{H}_{\rm aQPE}^{\rm XY} =\hat{a}^\dagger \hat{a} \otimes \hat{H}^{\rm XY}_{\rm target}$, 
where the cavity serves as a control register and the right part schematizes the coupling pattern for the target Hamiltonian $\hat{H}_{\rm XY}$.}
\label{fig:XYLayout}
\end{center}
\end{figure}

We develop a Hamiltonian engineering scheme to realize analog quantum phase estimation with the XY model as the target Hamiltonian, as illustrated in Fig.~\ref{fig:XYLayout}. The physical architecture consists of a control cavity dispersively coupled to target qubits through auxiliary coupler qubits, with the full Hamiltonian given by
\begin{align}
&\hat{H}_{\rm full}= 
\hbar \omega_c\, \hat{a}^\dagger \hat{a} 
+ \sum_{\mu} \frac{1}{2} \hbar \omega_{q\mu}\, \hat{\sigma}_\mu^{z} + \sum_{k} \hbar J_{\ell_k} \left(
\hat{\sigma}_{\ell_k}^{+} \hat{\sigma}_{k}^{-}
+ \hat{\sigma}_{\ell_k}^{-} \hat{\sigma}_{k}^{+}
\right) \notag\\
&+ \sum_{\langle k,k'\rangle} \hbar J_{m_{\langle k,k'\rangle}} \left[
\hat{\sigma}_{m_{\langle k,k'\rangle}}^{+} (\hat{\sigma}_{k}^{-} + \hat{\sigma}_{k'}^{-}) + \hat{\sigma}_{m_{\langle k,k'\rangle}}^{-} (\hat{\sigma}_{k}^{+} + \hat{\sigma}_{k'}^{+})
\right] \notag\\
&+ \sum_{\ell} \hbar g \left( \hat{a}^\dagger \hat{\sigma}_\ell^{-} + \hat{a} \hat{\sigma}_\ell^{+} \right) 
+ \sum_{m} \hbar g \left( \hat{a}^\dagger \hat{\sigma}_m^{-} + \hat{a} \hat{\sigma}_m^{+} \right),
\label{eqn:H_full}
\end{align}
where $\omega_c$ is the frequency of the control cavity. The system includes three types of qubits: target qubits (labeled by $k$), local coupler qubits (labeled by $\ell$), and cross coupler qubits (labeled by $m$), with $\mu \in \{k,\ell,m\}$ indexing all qubits, each at frequency $\omega_{q\mu}$. Each target qubit $k$ is coupled to a single local coupler $\ell_k$ with strength $J_{\ell_k}$, and each pair of target qubits $\langle k,k'\rangle$ is coupled to a single cross coupler $m_{\langle k,k'\rangle}$ with identical strength $J_{m_{\langle k,k'\rangle}}$. All coupler qubits are coupled to the cavity mode by a uniform Jaynes–Cummings-type interaction with strength $g$. We assume that all couplings are real without loss of generality.

Under the assumption $\abs{\Delta_\mu} \gg \abs{J_{\ell/m}}\gg \abs{g}$, where $\Delta_\mu= \omega_{q\mu}-\omega_c$ is the detuning between qubit $\mu$ and the cavity, the designed architecture realizes an engineered dispersive interaction between the cavity and the target qubits for aQPE implementation. The engineered interaction takes the form~\cite{note1}:
\begin{align}
&\hat{H}^{\rm{eng}}_{\rm aQPE}
\approx \hat{a}^\dagger \hat{a} \otimes \hat{H}^{\rm{eng}}_{\rm target} \notag\\
&= \hat{a}^\dagger \hat{a} \otimes \left[
\sum_k \frac{\hbar \lambda_k^{\rm eng}}{2} \hat{\sigma}^z_k +
\sum_{\langle k,k'\rangle} \hbar \eta_{kk'}^{\rm eng}
\left( \hat{\sigma}^+_k \hat{\sigma}^-_{k'} + \hat{\sigma}^-_k \hat{\sigma}^+_{k'} \right)
\right], \notag\\
&\lambda_k^{\rm eng} =
-\frac{2 g^2}{\Delta^3} \bigg(
\frac{1}{4} J_{\ell_k}^2 +
\frac{1}{4} \sum_{k'} J_{m_{\langle k,k'\rangle}}^2 +
\sum_{k'} J_{\ell_k} J_{m_{\langle k,k'\rangle}} \notag\\
&+ \sum_{k'\neq k''} J_{m_{\langle k,k'\rangle}} J_{m_{\langle k,k''\rangle}}
\bigg),\quad\text{and}\quad
\eta_{kk'}^{\rm eng} = \frac{3 g^2}{2\Delta^3} J_{m_{\langle k,k'\rangle}}^2,
\label{eqn:Heng}
\end{align}
where we assume a uniform detuning $\Delta_k = \omega_{qk} - \omega_c = \Delta$ for all target qubits.
The engineered interaction can be used to estimate eigenenergies of spin models combining tunable XY interactions and local longitudinal fields.  By choosing the physical parameters such that $\lambda_k^{\rm eng}=0$ for all $k$, one can realize a pure XY target Hamiltonian.

As a representative example of the target Hamiltonian, we consider a minimal model given by
\begin{align}
\hat{H}_{\rm target}^{\rm XY(2)} = \hbar \eta \left( \hat{\sigma}_1^+ \hat{\sigma}_2^- + \hat{\sigma}_1^- \hat{\sigma}_2^+ \right).
    \label{eqn:H2QXY}
\end{align}
The eigenenergies of this Hamiltonian are $\{-\hbar\eta,\ 0,\ 0,\ +\hbar\eta\}$, including a twofold degeneracy at zero energy. The corresponding ideal aQPE Hamiltonian takes the form
\begin{align}
\hat{H}_{\rm aQPE}^{\rm XY(2)} = \hat{a}^\dagger \hat{a} \otimes \hat{H}_{\rm target}^{\rm XY(2)},
\label{eqn:HaQPE2QXY}
\end{align}
which generates a conditional rotation of the cavity state according to the eigenenergies of the two-qubit target Hamiltonian. This interaction can be approximately realized through our Hamiltonian engineering scheme using two target qubits, one cavity register, and three auxiliary coupler qubits, resulting in an engineered interaction $\hat{H}^{\rm XY(2),eng}_{\rm aQPE} \approx \hat{a}^\dagger \hat{a} \otimes \hat{H}_{\rm target}^{\rm XY(2)}$ following Eq.~\eqref{eqn:Heng}, up to higher-order corrections~\cite{note1}.

\begin{figure}[!htbp]
\begin{center}
\includegraphics[width=0.9\linewidth]{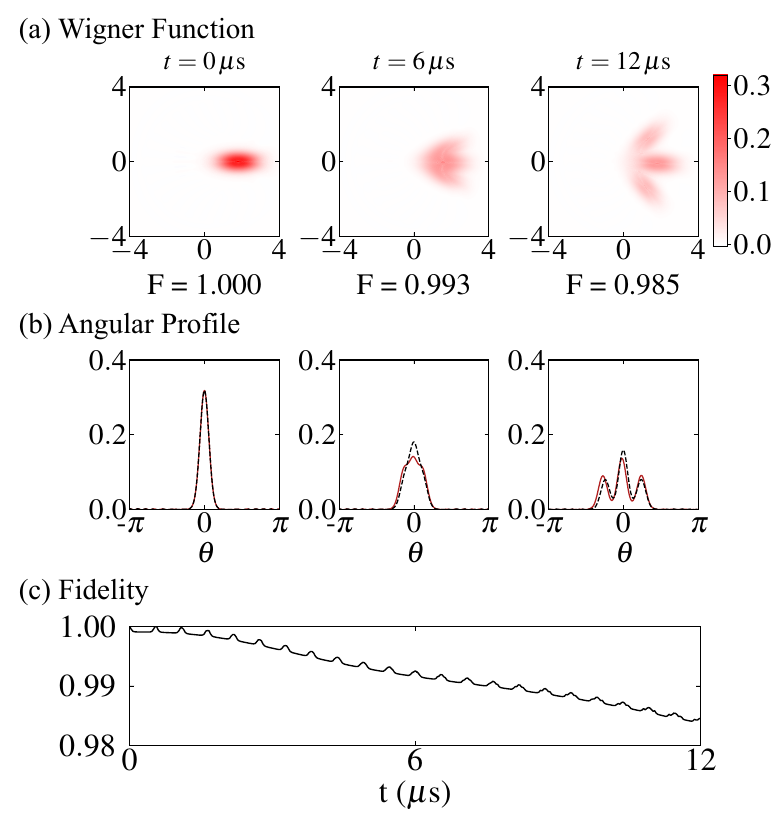}
\caption{Simulated dynamics of analog QPE for the two-qubit XY model with $\eta=10$ kHz under the engineered Hamiltonian. (a) Simulated cavity Wigner function snapshots at selected times. (b) Simulated angular profiles $ W(\beta e^{i\theta})$ sampled at the same time points as in (a). Dashed lines indicate the referenced evolution under the ideal aQPE Hamiltonian $\hat{H}_{\rm aQPE}^{\rm XY(2)}$. (c) Fidelity $F(t)$ between the cavity states under the engineered and ideal aQPE Hamiltonians. The cavity is initialized in $\ket{\beta=1.8,\xi=-0.4}$, while the target register is prepared in a uniform superposition of the target Hamiltonian eigenstates, $\ket{\psi}_q = \frac{1}{2} \left( \ket{e_1} + \ket{e_2} + \ket{e_3} + \ket{e_4} \right)$. The simulations are performed using QuTiP~\cite{Johansson2012,Johansson2013}, with parameters relevant for circuit QED devices~\cite{note3}.}
\label{fig:2QXYSim}
\end{center}
\end{figure}

We simulate the cavity evolution under the engineered aQPE interaction $\hat{H}^{\rm XY(2),eng}_{\rm aQPE}$,  as shown in Fig.~\ref{fig:2QXYSim}.  The simulation uses parameters compatible with current circuit QED devices~\cite{Paik2011,Nguyen2019,Krantz2019,Blais2021} to assess the spectral encoding performance and physical feasibility of the proposed scheme.  Starting from an initial squeezed cavity state and preparing the target register in an equal superposition of all eigenstates of $\hat{H}_{\rm target}^{\rm XY(2)}$, the resulting cavity evolution exhibits phase-space rotations that reflect the underlying eigenenergy structure.

Figure~\ref{fig:2QXYSim}(a) presents the simulated Wigner functions of the control cavity, which develop a clear threefold angular structure as the state accumulates conditional phase rotations governed by three distinct eigenenergies of $\hat{H}_{\rm target}^{\rm XY(2)}$. The corresponding angular profiles $W(\beta e^{i\theta})$ in Fig.~\ref{fig:2QXYSim}(b) reveal three emerging peaks, with the central peak carrying doubled amplitude due to spectral degeneracy. The fidelity $F(t)$ between the engineered and ideal aQPE dynamics~\cite{note2}, shown in Fig.~\ref{fig:2QXYSim}(c), remains above 98\% throughout the simulated timescale despite gradual reductions from higher-order corrections. These simulation results support the feasibility of implementing the aQPE protocol with reliable performance under realistic circuit QED conditions.

\section{Discussion}\label{sec:Discussion}
We have presented an analog quantum phase estimation (aQPE) protocol that encodes the eigenenergies of a target Hamiltonian as conditional phase-space rotations of a single-mode cavity. By implementing a continuous dispersive evolution and performing cavity phase-space tomography, the protocol directly extracts spectral information without relying on Trotterization, controlled unitaries, or other complex circuit-based operations. In contrast to conventional analog simulators where readout of eigenvalues remains challenging, the phase-space rotation in our protocol offers a directly measurable signature of the underlying Hamiltonian spectrum. We also provide a concrete Hamiltonian engineering method for implementing aQPE for the XY model, whose ground-state energy estimation is known to be QMA-complete, highlighting the applicability of aQPE to models that exhibit computational hardness.

Looking forward, the aQPE framework can be extended to a broader class of eigenvalue problems by engineering suitable dispersive interactions. Key applications include many-body physics and quantum algorithms such as Shor’s factoring and order-finding, both known to yield exponential quantum advantage. As the problem size increases, spectral crowding within the finite $2\pi$ rotation window will become relevant. The aQPE protocol can be further enhanced by incorporating techniques such as adiabatic algorithms~\cite{Farhi2001} to prepare low-energy eigenstates, or by designing filtering or postselection strategies to isolate specific eigencomponents. This protocol is readily implementable on experimental platforms that support tunable dispersive interactions, including but not limited to circuit QED systems, and offers a practical route toward scalable quantum phase estimation.

\begin{acknowledgments}
We thank J. Ignacio Cirac for helpful discussions.
W.-C.~Lin acknowledges funding support from the National Science and Technology Council, Taiwan, under grant \mbox{No.~113-2813-C-002-010-M,} and travel support from the Ministry of Education, Taiwan.  C.-H.~Wang acknowledges funding support from the National Science and Technology Council, Taiwan, under grants  \mbox{No.~111-2112-M-002-049-MY3,} \mbox{No.~113-2119-M-007-013-,} \mbox{No.~114-2119-M-007-013-,} and \mbox{No.~114-2124-M-002-003-,} and from the Office of Research and Development, National Taiwan University, under grants \mbox{No.~114L895001.}  C.-H. Wang is also grateful for the support from the Fubon Foundation, the Physics Division, National Center for Theoretical Sciences, Taiwan, and the Department of Physics, College of Science, National Taiwan University.
\end{acknowledgments}

\footnotetext{Detailed derivations will be provided in the Supplementary Information (in preparation)\label{note1}}
\footnotetext{Defined as respectively\label{note2}}
\footnotetext{Comprehensive simulation details will be provided in the Supplementary Information (in preparation)\label{note3}}

\bibliographystyle{apsrev4-2}
\bibliography{aQPE}

\begin{thebibliography}{60}%
\makeatletter
\providecommand \@ifxundefined [1]{%
 \@ifx{#1\undefined}
}%
\providecommand \@ifnum [1]{%
 \ifnum #1\expandafter \@firstoftwo
 \else \expandafter \@secondoftwo
 \fi
}%
\providecommand \@ifx [1]{%
 \ifx #1\expandafter \@firstoftwo
 \else \expandafter \@secondoftwo
 \fi
}%
\providecommand \natexlab [1]{#1}%
\providecommand \enquote  [1]{``#1''}%
\providecommand \bibnamefont  [1]{#1}%
\providecommand \bibfnamefont [1]{#1}%
\providecommand \citenamefont [1]{#1}%
\providecommand \href@noop [0]{\@secondoftwo}%
\providecommand \href [0]{\begingroup \@sanitize@url \@href}%
\providecommand \@href[1]{\@@startlink{#1}\@@href}%
\providecommand \@@href[1]{\endgroup#1\@@endlink}%
\providecommand \@sanitize@url [0]{\catcode `\\12\catcode `\$12\catcode
  `\&12\catcode `\#12\catcode `\^12\catcode `\_12\catcode `\%12\relax}%
\providecommand \@@startlink[1]{}%
\providecommand \@@endlink[0]{}%
\providecommand \url  [0]{\begingroup\@sanitize@url \@url }%
\providecommand \@url [1]{\endgroup\@href {#1}{\urlprefix }}%
\providecommand \urlprefix  [0]{URL }%
\providecommand \Eprint [0]{\href }%
\providecommand \doibase [0]{https://doi.org/}%
\providecommand \selectlanguage [0]{\@gobble}%
\providecommand \bibinfo  [0]{\@secondoftwo}%
\providecommand \bibfield  [0]{\@secondoftwo}%
\providecommand \translation [1]{[#1]}%
\providecommand \BibitemOpen [0]{}%
\providecommand \bibitemStop [0]{}%
\providecommand \bibitemNoStop [0]{.\EOS\space}%
\providecommand \EOS [0]{\spacefactor3000\relax}%
\providecommand \BibitemShut  [1]{\csname bibitem#1\endcsname}%
\let\auto@bib@innerbib\@empty
\bibitem [{\citenamefont {Lanyon}\ \emph {et~al.}(2010)\citenamefont {Lanyon},
  \citenamefont {Whitfield}, \citenamefont {Gillett}, \citenamefont {Goggin},
  \citenamefont {Almeida}, \citenamefont {Kassal}, \citenamefont {Biamonte},
  \citenamefont {Mohseni}, \citenamefont {Powell}, \citenamefont {Barbieri},
  \citenamefont {Aspuru-Guzik},\ and\ \citenamefont {White}}]{Lanyon2010}%
  \BibitemOpen
  \bibfield  {author} {\bibinfo {author} {\bibfnamefont {B.~P.}\ \bibnamefont
  {Lanyon}}, \bibinfo {author} {\bibfnamefont {J.~D.}\ \bibnamefont
  {Whitfield}}, \bibinfo {author} {\bibfnamefont {G.~G.}\ \bibnamefont
  {Gillett}}, \bibinfo {author} {\bibfnamefont {M.~E.}\ \bibnamefont {Goggin}},
  \bibinfo {author} {\bibfnamefont {M.~P.}\ \bibnamefont {Almeida}}, \bibinfo
  {author} {\bibfnamefont {I.}~\bibnamefont {Kassal}}, \bibinfo {author}
  {\bibfnamefont {J.~D.}\ \bibnamefont {Biamonte}}, \bibinfo {author}
  {\bibfnamefont {M.}~\bibnamefont {Mohseni}}, \bibinfo {author} {\bibfnamefont
  {B.~J.}\ \bibnamefont {Powell}}, \bibinfo {author} {\bibfnamefont
  {M.}~\bibnamefont {Barbieri}}, \bibinfo {author} {\bibfnamefont
  {A.}~\bibnamefont {Aspuru-Guzik}},\ and\ \bibinfo {author} {\bibfnamefont
  {A.~G.}\ \bibnamefont {White}},\ }\href {https://doi.org/10.1038/nchem.483}
  {\bibfield  {journal} {\bibinfo  {journal} {Nat. Chem.}\ }\textbf {\bibinfo
  {volume} {2}},\ \bibinfo {pages} {106} (\bibinfo {year} {2010})}\BibitemShut
  {NoStop}%
\bibitem [{\citenamefont {Cao}\ \emph {et~al.}(2019)\citenamefont {Cao},
  \citenamefont {Romero}, \citenamefont {Olson}, \citenamefont {Degroote},
  \citenamefont {Johnson}, \citenamefont {Kieferov{\'{a}}}, \citenamefont
  {Kivlichan}, \citenamefont {Menke}, \citenamefont {Peropadre}, \citenamefont
  {Sawaya}, \citenamefont {Sim}, \citenamefont {Veis},\ and\ \citenamefont
  {Aspuru-Guzik}}]{Cao2019}%
  \BibitemOpen
  \bibfield  {author} {\bibinfo {author} {\bibfnamefont {Y.}~\bibnamefont
  {Cao}}, \bibinfo {author} {\bibfnamefont {J.}~\bibnamefont {Romero}},
  \bibinfo {author} {\bibfnamefont {J.~P.}\ \bibnamefont {Olson}}, \bibinfo
  {author} {\bibfnamefont {M.}~\bibnamefont {Degroote}}, \bibinfo {author}
  {\bibfnamefont {P.~D.}\ \bibnamefont {Johnson}}, \bibinfo {author}
  {\bibfnamefont {M.}~\bibnamefont {Kieferov{\'{a}}}}, \bibinfo {author}
  {\bibfnamefont {I.~D.}\ \bibnamefont {Kivlichan}}, \bibinfo {author}
  {\bibfnamefont {T.}~\bibnamefont {Menke}}, \bibinfo {author} {\bibfnamefont
  {B.}~\bibnamefont {Peropadre}}, \bibinfo {author} {\bibfnamefont {N.~P.}\
  \bibnamefont {Sawaya}}, \bibinfo {author} {\bibfnamefont {S.}~\bibnamefont
  {Sim}}, \bibinfo {author} {\bibfnamefont {L.}~\bibnamefont {Veis}},\ and\
  \bibinfo {author} {\bibfnamefont {A.}~\bibnamefont {Aspuru-Guzik}},\ }\href
  {https://doi.org/10.1021/acs.chemrev.8b00803} {\bibfield  {journal} {\bibinfo
   {journal} {Chem. Rev.}\ }\textbf {\bibinfo {volume} {119}},\ \bibinfo
  {pages} {10856} (\bibinfo {year} {2019})}\BibitemShut {NoStop}%
\bibitem [{\citenamefont {McArdle}\ \emph {et~al.}(2020)\citenamefont
  {McArdle}, \citenamefont {Endo}, \citenamefont {Aspuru-Guzik}, \citenamefont
  {Benjamin},\ and\ \citenamefont {Yuan}}]{McArdle2020}%
  \BibitemOpen
  \bibfield  {author} {\bibinfo {author} {\bibfnamefont {S.}~\bibnamefont
  {McArdle}}, \bibinfo {author} {\bibfnamefont {S.}~\bibnamefont {Endo}},
  \bibinfo {author} {\bibfnamefont {A.}~\bibnamefont {Aspuru-Guzik}}, \bibinfo
  {author} {\bibfnamefont {S.~C.}\ \bibnamefont {Benjamin}},\ and\ \bibinfo
  {author} {\bibfnamefont {X.}~\bibnamefont {Yuan}},\ }\href
  {https://doi.org/10.1103/RevModPhys.92.015003} {\bibfield  {journal}
  {\bibinfo  {journal} {Rev. Mod. Phys.}\ }\textbf {\bibinfo {volume} {92}},\
  \bibinfo {pages} {015003} (\bibinfo {year} {2020})}\BibitemShut {NoStop}%
\bibitem [{\citenamefont {Wecker}\ \emph {et~al.}(2015)\citenamefont {Wecker},
  \citenamefont {Hastings}, \citenamefont {Wiebe}, \citenamefont {Clark},
  \citenamefont {Nayak},\ and\ \citenamefont {Troyer}}]{Wecker2015}%
  \BibitemOpen
  \bibfield  {author} {\bibinfo {author} {\bibfnamefont {D.}~\bibnamefont
  {Wecker}}, \bibinfo {author} {\bibfnamefont {M.~B.}\ \bibnamefont
  {Hastings}}, \bibinfo {author} {\bibfnamefont {N.}~\bibnamefont {Wiebe}},
  \bibinfo {author} {\bibfnamefont {B.~K.}\ \bibnamefont {Clark}}, \bibinfo
  {author} {\bibfnamefont {C.}~\bibnamefont {Nayak}},\ and\ \bibinfo {author}
  {\bibfnamefont {M.}~\bibnamefont {Troyer}},\ }\href
  {https://doi.org/10.1103/PhysRevA.92.062318} {\bibfield  {journal} {\bibinfo
  {journal} {Phys. Rev. A}\ }\textbf {\bibinfo {volume} {92}},\ \bibinfo
  {pages} {062318} (\bibinfo {year} {2015})}\BibitemShut {NoStop}%
\bibitem [{\citenamefont {Babbush}\ \emph {et~al.}(2018)\citenamefont
  {Babbush}, \citenamefont {Wiebe}, \citenamefont {McClean}, \citenamefont
  {McClain}, \citenamefont {Neven},\ and\ \citenamefont {Chan}}]{Babbush2018}%
  \BibitemOpen
  \bibfield  {author} {\bibinfo {author} {\bibfnamefont {R.}~\bibnamefont
  {Babbush}}, \bibinfo {author} {\bibfnamefont {N.}~\bibnamefont {Wiebe}},
  \bibinfo {author} {\bibfnamefont {J.}~\bibnamefont {McClean}}, \bibinfo
  {author} {\bibfnamefont {J.}~\bibnamefont {McClain}}, \bibinfo {author}
  {\bibfnamefont {H.}~\bibnamefont {Neven}},\ and\ \bibinfo {author}
  {\bibfnamefont {G.~K.~L.}\ \bibnamefont {Chan}},\ }\href
  {https://doi.org/10.1103/PhysRevX.8.011044} {\bibfield  {journal} {\bibinfo
  {journal} {Phys. Rev. X}\ }\textbf {\bibinfo {volume} {8}},\ \bibinfo {pages}
  {11044} (\bibinfo {year} {2018})}\BibitemShut {NoStop}%
\bibitem [{\citenamefont {Bauer}\ \emph {et~al.}(2020)\citenamefont {Bauer},
  \citenamefont {Bravyi}, \citenamefont {Motta},\ and\ \citenamefont {{Kin-Lic
  Chan}}}]{Bauer2020}%
  \BibitemOpen
  \bibfield  {author} {\bibinfo {author} {\bibfnamefont {B.}~\bibnamefont
  {Bauer}}, \bibinfo {author} {\bibfnamefont {S.}~\bibnamefont {Bravyi}},
  \bibinfo {author} {\bibfnamefont {M.}~\bibnamefont {Motta}},\ and\ \bibinfo
  {author} {\bibfnamefont {G.}~\bibnamefont {{Kin-Lic Chan}}},\ }\href
  {https://doi.org/10.1021/acs.chemrev.9b00829} {\bibfield  {journal} {\bibinfo
   {journal} {Chem. Rev.}\ }\textbf {\bibinfo {volume} {120}},\ \bibinfo
  {pages} {12685} (\bibinfo {year} {2020})}\BibitemShut {NoStop}%
\bibitem [{\citenamefont {Sachdev}(2000)}]{Sachdev2000}%
  \BibitemOpen
  \bibfield  {author} {\bibinfo {author} {\bibfnamefont {S.}~\bibnamefont
  {Sachdev}},\ }\href@noop {} {\emph {\bibinfo {title} {Quantum Phase
  Transitions}}}\ (\bibinfo  {publisher} {Cambridge University Press,
  Cambridge, UK},\ \bibinfo {year} {2000})\BibitemShut {NoStop}%
\bibitem [{\citenamefont {Roushan}\ \emph {et~al.}(2017)\citenamefont
  {Roushan}, \citenamefont {Neill}, \citenamefont {Tangpanitanon},
  \citenamefont {Bastidas}, \citenamefont {Megrant}, \citenamefont {Barends},
  \citenamefont {Chen}, \citenamefont {Chen}, \citenamefont {Chiaro},
  \citenamefont {Dunsworth}, \citenamefont {Fowler}, \citenamefont {Foxen},
  \citenamefont {Giustina}, \citenamefont {Jeffrey}, \citenamefont {Kelly},
  \citenamefont {Lucero}, \citenamefont {Mutus}, \citenamefont {Neeley},
  \citenamefont {Quintana}, \citenamefont {Sank}, \citenamefont {Vainsencher},
  \citenamefont {Wenner}, \citenamefont {White}, \citenamefont {Neven},
  \citenamefont {Angelakis},\ and\ \citenamefont {Martinis}}]{Roushan2017}%
  \BibitemOpen
  \bibfield  {author} {\bibinfo {author} {\bibfnamefont {P.}~\bibnamefont
  {Roushan}}, \bibinfo {author} {\bibfnamefont {C.}~\bibnamefont {Neill}},
  \bibinfo {author} {\bibfnamefont {J.}~\bibnamefont {Tangpanitanon}}, \bibinfo
  {author} {\bibfnamefont {V.~M.}\ \bibnamefont {Bastidas}}, \bibinfo {author}
  {\bibfnamefont {A.}~\bibnamefont {Megrant}}, \bibinfo {author} {\bibfnamefont
  {R.}~\bibnamefont {Barends}}, \bibinfo {author} {\bibfnamefont
  {Y.}~\bibnamefont {Chen}}, \bibinfo {author} {\bibfnamefont {Z.}~\bibnamefont
  {Chen}}, \bibinfo {author} {\bibfnamefont {B.}~\bibnamefont {Chiaro}},
  \bibinfo {author} {\bibfnamefont {A.}~\bibnamefont {Dunsworth}}, \bibinfo
  {author} {\bibfnamefont {A.}~\bibnamefont {Fowler}}, \bibinfo {author}
  {\bibfnamefont {B.}~\bibnamefont {Foxen}}, \bibinfo {author} {\bibfnamefont
  {M.}~\bibnamefont {Giustina}}, \bibinfo {author} {\bibfnamefont
  {E.}~\bibnamefont {Jeffrey}}, \bibinfo {author} {\bibfnamefont
  {J.}~\bibnamefont {Kelly}}, \bibinfo {author} {\bibfnamefont
  {E.}~\bibnamefont {Lucero}}, \bibinfo {author} {\bibfnamefont
  {J.}~\bibnamefont {Mutus}}, \bibinfo {author} {\bibfnamefont
  {M.}~\bibnamefont {Neeley}}, \bibinfo {author} {\bibfnamefont
  {C.}~\bibnamefont {Quintana}}, \bibinfo {author} {\bibfnamefont
  {D.}~\bibnamefont {Sank}}, \bibinfo {author} {\bibfnamefont {A.}~\bibnamefont
  {Vainsencher}}, \bibinfo {author} {\bibfnamefont {J.}~\bibnamefont {Wenner}},
  \bibinfo {author} {\bibfnamefont {T.}~\bibnamefont {White}}, \bibinfo
  {author} {\bibfnamefont {H.}~\bibnamefont {Neven}}, \bibinfo {author}
  {\bibfnamefont {D.~G.}\ \bibnamefont {Angelakis}},\ and\ \bibinfo {author}
  {\bibfnamefont {J.}~\bibnamefont {Martinis}},\ }\href
  {https://doi.org/10.1126/science.aao1401} {\bibfield  {journal} {\bibinfo
  {journal} {Science}\ }\textbf {\bibinfo {volume} {358}},\ \bibinfo {pages}
  {1175} (\bibinfo {year} {2017})}\BibitemShut {NoStop}%
\bibitem [{\citenamefont {Biamonte}\ \emph {et~al.}(2017)\citenamefont
  {Biamonte}, \citenamefont {Wittek}, \citenamefont {Pancotti}, \citenamefont
  {Rebentrost}, \citenamefont {Wiebe},\ and\ \citenamefont
  {Lloyd}}]{Biamonte2017}%
  \BibitemOpen
  \bibfield  {author} {\bibinfo {author} {\bibfnamefont {J.}~\bibnamefont
  {Biamonte}}, \bibinfo {author} {\bibfnamefont {P.}~\bibnamefont {Wittek}},
  \bibinfo {author} {\bibfnamefont {N.}~\bibnamefont {Pancotti}}, \bibinfo
  {author} {\bibfnamefont {P.}~\bibnamefont {Rebentrost}}, \bibinfo {author}
  {\bibfnamefont {N.}~\bibnamefont {Wiebe}},\ and\ \bibinfo {author}
  {\bibfnamefont {S.}~\bibnamefont {Lloyd}},\ }\href
  {https://doi.org/10.1038/nature23474} {\bibfield  {journal} {\bibinfo
  {journal} {Nature}\ }\textbf {\bibinfo {volume} {549}},\ \bibinfo {pages}
  {195} (\bibinfo {year} {2017})}\BibitemShut {NoStop}%
\bibitem [{\citenamefont {Coleman}(1963)}]{Coleman1963}%
  \BibitemOpen
  \bibfield  {author} {\bibinfo {author} {\bibfnamefont {A.~J.}\ \bibnamefont
  {Coleman}},\ }\href {https://doi.org/10.1103/RevModPhys.35.668} {\bibfield
  {journal} {\bibinfo  {journal} {Rev. Mod. Phys.}\ }\textbf {\bibinfo {volume}
  {35}},\ \bibinfo {pages} {668} (\bibinfo {year} {1963})}\BibitemShut
  {NoStop}%
\bibitem [{\citenamefont {Liu}\ \emph {et~al.}(2007)\citenamefont {Liu},
  \citenamefont {Christandl},\ and\ \citenamefont {Verstraete}}]{Liu2007}%
  \BibitemOpen
  \bibfield  {author} {\bibinfo {author} {\bibfnamefont {Y.~K.}\ \bibnamefont
  {Liu}}, \bibinfo {author} {\bibfnamefont {M.}~\bibnamefont {Christandl}},\
  and\ \bibinfo {author} {\bibfnamefont {F.}~\bibnamefont {Verstraete}},\
  }\href {https://doi.org/10.1103/PhysRevLett.98.110503} {\bibfield  {journal}
  {\bibinfo  {journal} {Phys. Rev. Lett.}\ }\textbf {\bibinfo {volume} {98}},\
  \bibinfo {pages} {110503} (\bibinfo {year} {2007})}\BibitemShut {NoStop}%
\bibitem [{\citenamefont {Schuch}\ and\ \citenamefont
  {Verstraete}(2009)}]{Schuch2009}%
  \BibitemOpen
  \bibfield  {author} {\bibinfo {author} {\bibfnamefont {N.}~\bibnamefont
  {Schuch}}\ and\ \bibinfo {author} {\bibfnamefont {F.}~\bibnamefont
  {Verstraete}},\ }\href {https://doi.org/10.1038/nphys1370} {\bibfield
  {journal} {\bibinfo  {journal} {Nat. Phys.}\ }\textbf {\bibinfo {volume}
  {5}},\ \bibinfo {pages} {732} (\bibinfo {year} {2009})}\BibitemShut {NoStop}%
\bibitem [{\citenamefont {Jones}(2015)}]{Jones2015}%
  \BibitemOpen
  \bibfield  {author} {\bibinfo {author} {\bibfnamefont {R.~O.}\ \bibnamefont
  {Jones}},\ }\href {https://doi.org/10.1103/RevModPhys.87.897} {\bibfield
  {journal} {\bibinfo  {journal} {Rev. Mod. Phys.}\ }\textbf {\bibinfo {volume}
  {87}},\ \bibinfo {pages} {897} (\bibinfo {year} {2015})}\BibitemShut
  {NoStop}%
\bibitem [{\citenamefont {Foulkes}\ \emph {et~al.}(2001)\citenamefont
  {Foulkes}, \citenamefont {Mitas}, \citenamefont {Needs},\ and\ \citenamefont
  {Rajagopal}}]{Foulkes2001}%
  \BibitemOpen
  \bibfield  {author} {\bibinfo {author} {\bibfnamefont {W.~M.}\ \bibnamefont
  {Foulkes}}, \bibinfo {author} {\bibfnamefont {L.}~\bibnamefont {Mitas}},
  \bibinfo {author} {\bibfnamefont {R.~J.}\ \bibnamefont {Needs}},\ and\
  \bibinfo {author} {\bibfnamefont {G.}~\bibnamefont {Rajagopal}},\ }\href
  {https://doi.org/10.1103/RevModPhys.73.33} {\bibfield  {journal} {\bibinfo
  {journal} {Rev. Mod. Phys.}\ }\textbf {\bibinfo {volume} {73}},\ \bibinfo
  {pages} {33} (\bibinfo {year} {2001})}\BibitemShut {NoStop}%
\bibitem [{\citenamefont {L{\"{u}}chow}(2011)}]{Luchow2011}%
  \BibitemOpen
  \bibfield  {author} {\bibinfo {author} {\bibfnamefont {A.}~\bibnamefont
  {L{\"{u}}chow}},\ }\href {https://doi.org/10.1002/wcms.40} {\bibfield
  {journal} {\bibinfo  {journal} {Wiley Interdiscip. Rev. Comput. Mol. Sci.}\
  }\textbf {\bibinfo {volume} {1}},\ \bibinfo {pages} {388} (\bibinfo {year}
  {2011})}\BibitemShut {NoStop}%
\bibitem [{\citenamefont {Carlson}\ \emph {et~al.}(2015)\citenamefont
  {Carlson}, \citenamefont {Gandolfi}, \citenamefont {Pederiva}, \citenamefont
  {Pieper}, \citenamefont {Schiavilla}, \citenamefont {Schmidt},\ and\
  \citenamefont {Wiringa}}]{Carlson2015}%
  \BibitemOpen
  \bibfield  {author} {\bibinfo {author} {\bibfnamefont {J.}~\bibnamefont
  {Carlson}}, \bibinfo {author} {\bibfnamefont {S.}~\bibnamefont {Gandolfi}},
  \bibinfo {author} {\bibfnamefont {F.}~\bibnamefont {Pederiva}}, \bibinfo
  {author} {\bibfnamefont {S.~C.}\ \bibnamefont {Pieper}}, \bibinfo {author}
  {\bibfnamefont {R.}~\bibnamefont {Schiavilla}}, \bibinfo {author}
  {\bibfnamefont {K.~E.}\ \bibnamefont {Schmidt}},\ and\ \bibinfo {author}
  {\bibfnamefont {R.~B.}\ \bibnamefont {Wiringa}},\ }\href
  {https://doi.org/10.1103/RevModPhys.87.1067} {\bibfield  {journal} {\bibinfo
  {journal} {Rev. Mod. Phys.}\ }\textbf {\bibinfo {volume} {87}},\ \bibinfo
  {pages} {1067} (\bibinfo {year} {2015})}\BibitemShut {NoStop}%
\bibitem [{\citenamefont {Kitaev}(1995)}]{Kitaev1995}%
  \BibitemOpen
  \bibfield  {author} {\bibinfo {author} {\bibfnamefont {A.~Y.}\ \bibnamefont
  {Kitaev}},\ }\href@noop {} {\bibinfo {title} {{Quantum measurements and the
  Abelian Stabilizer Problem}}} (\bibinfo {year} {1995}),\ \Eprint
  {https://arxiv.org/abs/arXiv:quant-ph/9511026} {arXiv:quant-ph/9511026}
  \BibitemShut {NoStop}%
\bibitem [{\citenamefont {Cleve}\ \emph {et~al.}(1998)\citenamefont {Cleve},
  \citenamefont {Ekert}, \citenamefont {Macchiavello},\ and\ \citenamefont
  {Mosca}}]{Cleve1998}%
  \BibitemOpen
  \bibfield  {author} {\bibinfo {author} {\bibfnamefont {R.}~\bibnamefont
  {Cleve}}, \bibinfo {author} {\bibfnamefont {A.}~\bibnamefont {Ekert}},
  \bibinfo {author} {\bibfnamefont {C.}~\bibnamefont {Macchiavello}},\ and\
  \bibinfo {author} {\bibfnamefont {M.}~\bibnamefont {Mosca}},\ }\href
  {https://doi.org/10.1098/rspa.1998.0164} {\bibfield  {journal} {\bibinfo
  {journal} {Proc. R. Soc. A Math. Phys. Eng. Sci.}\ }\textbf {\bibinfo
  {volume} {454}},\ \bibinfo {pages} {339} (\bibinfo {year}
  {1998})}\BibitemShut {NoStop}%
\bibitem [{\citenamefont {Abrams}\ and\ \citenamefont
  {Lloyd}(1999)}]{Abrams1999}%
  \BibitemOpen
  \bibfield  {author} {\bibinfo {author} {\bibfnamefont {D.~S.}\ \bibnamefont
  {Abrams}}\ and\ \bibinfo {author} {\bibfnamefont {S.}~\bibnamefont {Lloyd}},\
  }\href {https://doi.org/10.1103/PhysRevLett.83.5162} {\bibfield  {journal}
  {\bibinfo  {journal} {Phys. Rev. Lett.}\ }\textbf {\bibinfo {volume} {83}},\
  \bibinfo {pages} {5162} (\bibinfo {year} {1999})}\BibitemShut {NoStop}%
\bibitem [{\citenamefont {Aspuru-Guzik}\ \emph {et~al.}(2005)\citenamefont
  {Aspuru-Guzik}, \citenamefont {Dutoi}, \citenamefont {Love},\ and\
  \citenamefont {Head-Gordon}}]{Aspuru2005}%
  \BibitemOpen
  \bibfield  {author} {\bibinfo {author} {\bibfnamefont {A.}~\bibnamefont
  {Aspuru-Guzik}}, \bibinfo {author} {\bibfnamefont {A.~D.}\ \bibnamefont
  {Dutoi}}, \bibinfo {author} {\bibfnamefont {P.~J.}\ \bibnamefont {Love}},\
  and\ \bibinfo {author} {\bibfnamefont {M.}~\bibnamefont {Head-Gordon}},\
  }\href {https://doi.org/10.1126/science.1113479} {\bibfield  {journal}
  {\bibinfo  {journal} {Science}\ }\textbf {\bibinfo {volume} {309}},\ \bibinfo
  {pages} {1704} (\bibinfo {year} {2005})}\BibitemShut {NoStop}%
\bibitem [{\citenamefont {Shor}(1999)}]{Shor1999}%
  \BibitemOpen
  \bibfield  {author} {\bibinfo {author} {\bibfnamefont {P.~W.}\ \bibnamefont
  {Shor}},\ }\href {https://doi.org/10.1137/S0036144598347011} {\bibfield
  {journal} {\bibinfo  {journal} {SIAM J. Comput.}\ }\textbf {\bibinfo {volume}
  {41}},\ \bibinfo {pages} {303} (\bibinfo {year} {1999})}\BibitemShut
  {NoStop}%
\bibitem [{\citenamefont {Trotter}(1959)}]{Trotter1959}%
  \BibitemOpen
  \bibfield  {author} {\bibinfo {author} {\bibfnamefont {H.~F.}\ \bibnamefont
  {Trotter}},\ }\href {https://doi.org/10.1090/S0002-9939-1959-0108732-6}
  {\bibfield  {journal} {\bibinfo  {journal} {Proc. Am. Math. Soc.}\ }\textbf
  {\bibinfo {volume} {10}},\ \bibinfo {pages} {545} (\bibinfo {year}
  {1959})}\BibitemShut {NoStop}%
\bibitem [{\citenamefont {Griffiths}\ and\ \citenamefont
  {Niu}(1996)}]{Griffiths1996}%
  \BibitemOpen
  \bibfield  {author} {\bibinfo {author} {\bibfnamefont {R.~B.}\ \bibnamefont
  {Griffiths}}\ and\ \bibinfo {author} {\bibfnamefont {C.~S.}\ \bibnamefont
  {Niu}},\ }\href {https://doi.org/10.1103/PhysRevLett.76.3228} {\bibfield
  {journal} {\bibinfo  {journal} {Phys. Rev. Lett.}\ }\textbf {\bibinfo
  {volume} {76}},\ \bibinfo {pages} {3228} (\bibinfo {year}
  {1996})}\BibitemShut {NoStop}%
\bibitem [{\citenamefont {Wiebe}\ and\ \citenamefont
  {Granade}(2016)}]{Wiebe2016}%
  \BibitemOpen
  \bibfield  {author} {\bibinfo {author} {\bibfnamefont {N.}~\bibnamefont
  {Wiebe}}\ and\ \bibinfo {author} {\bibfnamefont {C.}~\bibnamefont
  {Granade}},\ }\href {https://doi.org/10.1103/PhysRevLett.117.010503}
  {\bibfield  {journal} {\bibinfo  {journal} {Phys. Rev. Lett.}\ }\textbf
  {\bibinfo {volume} {117}},\ \bibinfo {pages} {010503} (\bibinfo {year}
  {2016})}\BibitemShut {NoStop}%
\bibitem [{\citenamefont {Russo}\ \emph {et~al.}(2021)\citenamefont {Russo},
  \citenamefont {Rudinger}, \citenamefont {Morrison},\ and\ \citenamefont
  {Baczewski}}]{Russo2021}%
  \BibitemOpen
  \bibfield  {author} {\bibinfo {author} {\bibfnamefont {A.~E.}\ \bibnamefont
  {Russo}}, \bibinfo {author} {\bibfnamefont {K.~M.}\ \bibnamefont {Rudinger}},
  \bibinfo {author} {\bibfnamefont {B.~C.}\ \bibnamefont {Morrison}},\ and\
  \bibinfo {author} {\bibfnamefont {A.~D.}\ \bibnamefont {Baczewski}},\ }\href
  {https://doi.org/10.1103/PhysRevLett.126.210501} {\bibfield  {journal}
  {\bibinfo  {journal} {Phys. Rev. Lett.}\ }\textbf {\bibinfo {volume} {126}},\
  \bibinfo {pages} {210501} (\bibinfo {year} {2021})}\BibitemShut {NoStop}%
\bibitem [{\citenamefont {Ni}\ \emph {et~al.}(2023)\citenamefont {Ni},
  \citenamefont {Li},\ and\ \citenamefont {Ying}}]{Ni2023}%
  \BibitemOpen
  \bibfield  {author} {\bibinfo {author} {\bibfnamefont {H.}~\bibnamefont
  {Ni}}, \bibinfo {author} {\bibfnamefont {H.}~\bibnamefont {Li}},\ and\
  \bibinfo {author} {\bibfnamefont {L.}~\bibnamefont {Ying}},\ }\href
  {https://doi.org/10.22331/q-2023-11-06-1165} {\bibfield  {journal} {\bibinfo
  {journal} {Quantum}\ }\textbf {\bibinfo {volume} {7}},\ \bibinfo {pages}
  {1165} (\bibinfo {year} {2023})}\BibitemShut {NoStop}%
\bibitem [{\citenamefont {Peruzzo}\ \emph {et~al.}(2014)\citenamefont
  {Peruzzo}, \citenamefont {McClean}, \citenamefont {Shadbolt}, \citenamefont
  {Yung}, \citenamefont {Zhou}, \citenamefont {Love}, \citenamefont
  {Aspuru-Guzik},\ and\ \citenamefont {O'Brien}}]{Peruzzo2014}%
  \BibitemOpen
  \bibfield  {author} {\bibinfo {author} {\bibfnamefont {A.}~\bibnamefont
  {Peruzzo}}, \bibinfo {author} {\bibfnamefont {J.}~\bibnamefont {McClean}},
  \bibinfo {author} {\bibfnamefont {P.}~\bibnamefont {Shadbolt}}, \bibinfo
  {author} {\bibfnamefont {M.~H.}\ \bibnamefont {Yung}}, \bibinfo {author}
  {\bibfnamefont {X.~Q.}\ \bibnamefont {Zhou}}, \bibinfo {author}
  {\bibfnamefont {P.~J.}\ \bibnamefont {Love}}, \bibinfo {author}
  {\bibfnamefont {A.}~\bibnamefont {Aspuru-Guzik}},\ and\ \bibinfo {author}
  {\bibfnamefont {J.~L.}\ \bibnamefont {O'Brien}},\ }\href
  {https://doi.org/10.1038/ncomms5213} {\bibfield  {journal} {\bibinfo
  {journal} {Nat. Commun.}\ }\textbf {\bibinfo {volume} {5}},\ \bibinfo {pages}
  {4213} (\bibinfo {year} {2014})}\BibitemShut {NoStop}%
\bibitem [{\citenamefont {McClean}\ \emph {et~al.}(2016)\citenamefont
  {McClean}, \citenamefont {Romero}, \citenamefont {Babbush},\ and\
  \citenamefont {Aspuru-Guzik}}]{McClean2016}%
  \BibitemOpen
  \bibfield  {author} {\bibinfo {author} {\bibfnamefont {J.~R.}\ \bibnamefont
  {McClean}}, \bibinfo {author} {\bibfnamefont {J.}~\bibnamefont {Romero}},
  \bibinfo {author} {\bibfnamefont {R.}~\bibnamefont {Babbush}},\ and\ \bibinfo
  {author} {\bibfnamefont {A.}~\bibnamefont {Aspuru-Guzik}},\ }\href
  {https://doi.org/10.1088/1367-2630/18/2/023023} {\bibfield  {journal}
  {\bibinfo  {journal} {New J. Phys.}\ }\textbf {\bibinfo {volume} {18}},\
  \bibinfo {pages} {023023} (\bibinfo {year} {2016})}\BibitemShut {NoStop}%
\bibitem [{\citenamefont {Tilly}\ \emph {et~al.}(2022)\citenamefont {Tilly},
  \citenamefont {Chen}, \citenamefont {Cao}, \citenamefont {Picozzi},
  \citenamefont {Setia}, \citenamefont {Li}, \citenamefont {Grant},
  \citenamefont {Wossnig}, \citenamefont {Rungger}, \citenamefont {Booth},\
  and\ \citenamefont {Tennyson}}]{Tilly2022}%
  \BibitemOpen
  \bibfield  {author} {\bibinfo {author} {\bibfnamefont {J.}~\bibnamefont
  {Tilly}}, \bibinfo {author} {\bibfnamefont {H.}~\bibnamefont {Chen}},
  \bibinfo {author} {\bibfnamefont {S.}~\bibnamefont {Cao}}, \bibinfo {author}
  {\bibfnamefont {D.}~\bibnamefont {Picozzi}}, \bibinfo {author} {\bibfnamefont
  {K.}~\bibnamefont {Setia}}, \bibinfo {author} {\bibfnamefont
  {Y.}~\bibnamefont {Li}}, \bibinfo {author} {\bibfnamefont {E.}~\bibnamefont
  {Grant}}, \bibinfo {author} {\bibfnamefont {L.}~\bibnamefont {Wossnig}},
  \bibinfo {author} {\bibfnamefont {I.}~\bibnamefont {Rungger}}, \bibinfo
  {author} {\bibfnamefont {G.~H.}\ \bibnamefont {Booth}},\ and\ \bibinfo
  {author} {\bibfnamefont {J.}~\bibnamefont {Tennyson}},\ }\href
  {https://doi.org/10.1016/j.physrep.2022.08.003} {\bibfield  {journal}
  {\bibinfo  {journal} {Phys. Rep.}\ }\textbf {\bibinfo {volume} {986}},\
  \bibinfo {pages} {1} (\bibinfo {year} {2022})}\BibitemShut {NoStop}%
\bibitem [{\citenamefont {Senko}\ \emph {et~al.}(2014)\citenamefont {Senko},
  \citenamefont {Smith}, \citenamefont {Richerme}, \citenamefont {Lee},
  \citenamefont {Campbell},\ and\ \citenamefont {Monroe}}]{Senko2014}%
  \BibitemOpen
  \bibfield  {author} {\bibinfo {author} {\bibfnamefont {C.}~\bibnamefont
  {Senko}}, \bibinfo {author} {\bibfnamefont {J.}~\bibnamefont {Smith}},
  \bibinfo {author} {\bibfnamefont {P.}~\bibnamefont {Richerme}}, \bibinfo
  {author} {\bibfnamefont {A.}~\bibnamefont {Lee}}, \bibinfo {author}
  {\bibfnamefont {W.~C.}\ \bibnamefont {Campbell}},\ and\ \bibinfo {author}
  {\bibfnamefont {C.}~\bibnamefont {Monroe}},\ }\href
  {https://doi.org/10.1126/science.1251422} {\bibfield  {journal} {\bibinfo
  {journal} {Science}\ }\textbf {\bibinfo {volume} {345}},\ \bibinfo {pages}
  {430} (\bibinfo {year} {2014})}\BibitemShut {NoStop}%
\bibitem [{\citenamefont {Hauke}\ \emph {et~al.}(2012)\citenamefont {Hauke},
  \citenamefont {Cucchietti}, \citenamefont {Tagliacozzo}, \citenamefont
  {Deutsch},\ and\ \citenamefont {Lewenstein}}]{Hauke2012}%
  \BibitemOpen
  \bibfield  {author} {\bibinfo {author} {\bibfnamefont {P.}~\bibnamefont
  {Hauke}}, \bibinfo {author} {\bibfnamefont {F.~M.}\ \bibnamefont
  {Cucchietti}}, \bibinfo {author} {\bibfnamefont {L.}~\bibnamefont
  {Tagliacozzo}}, \bibinfo {author} {\bibfnamefont {I.}~\bibnamefont
  {Deutsch}},\ and\ \bibinfo {author} {\bibfnamefont {M.}~\bibnamefont
  {Lewenstein}},\ }\href {https://doi.org/10.1088/0034-4885/75/8/082401}
  {\bibfield  {journal} {\bibinfo  {journal} {Reports Prog. Phys.}\ }\textbf
  {\bibinfo {volume} {75}},\ \bibinfo {pages} {082401} (\bibinfo {year}
  {2012})}\BibitemShut {NoStop}%
\bibitem [{\citenamefont {Wallraff}\ \emph {et~al.}(2005)\citenamefont
  {Wallraff}, \citenamefont {Schuster}, \citenamefont {Blais}, \citenamefont
  {Frunzio}, \citenamefont {Majer}, \citenamefont {Devoret}, \citenamefont
  {Girvin},\ and\ \citenamefont {Schoelkopf}}]{Wallraff2005}%
  \BibitemOpen
  \bibfield  {author} {\bibinfo {author} {\bibfnamefont {A.}~\bibnamefont
  {Wallraff}}, \bibinfo {author} {\bibfnamefont {D.~I.}\ \bibnamefont
  {Schuster}}, \bibinfo {author} {\bibfnamefont {A.}~\bibnamefont {Blais}},
  \bibinfo {author} {\bibfnamefont {L.}~\bibnamefont {Frunzio}}, \bibinfo
  {author} {\bibfnamefont {J.}~\bibnamefont {Majer}}, \bibinfo {author}
  {\bibfnamefont {M.~H.}\ \bibnamefont {Devoret}}, \bibinfo {author}
  {\bibfnamefont {S.~M.}\ \bibnamefont {Girvin}},\ and\ \bibinfo {author}
  {\bibfnamefont {R.~J.}\ \bibnamefont {Schoelkopf}},\ }\href
  {https://doi.org/10.1103/PhysRevLett.95.060501} {\bibfield  {journal}
  {\bibinfo  {journal} {Phys. Rev. Lett.}\ }\textbf {\bibinfo {volume} {95}},\
  \bibinfo {pages} {060501} (\bibinfo {year} {2005})}\BibitemShut {NoStop}%
\bibitem [{\citenamefont {Blais}\ \emph {et~al.}(2007)\citenamefont {Blais},
  \citenamefont {Gambetta}, \citenamefont {Wallraff}, \citenamefont {Schuster},
  \citenamefont {Girvin}, \citenamefont {Devoret},\ and\ \citenamefont
  {Schoelkopf}}]{Blais2007}%
  \BibitemOpen
  \bibfield  {author} {\bibinfo {author} {\bibfnamefont {A.}~\bibnamefont
  {Blais}}, \bibinfo {author} {\bibfnamefont {J.}~\bibnamefont {Gambetta}},
  \bibinfo {author} {\bibfnamefont {A.}~\bibnamefont {Wallraff}}, \bibinfo
  {author} {\bibfnamefont {D.~I.}\ \bibnamefont {Schuster}}, \bibinfo {author}
  {\bibfnamefont {S.~M.}\ \bibnamefont {Girvin}}, \bibinfo {author}
  {\bibfnamefont {M.~H.}\ \bibnamefont {Devoret}},\ and\ \bibinfo {author}
  {\bibfnamefont {R.~J.}\ \bibnamefont {Schoelkopf}},\ }\href
  {https://doi.org/10.1103/PhysRevA.75.032329} {\bibfield  {journal} {\bibinfo
  {journal} {Phys. Rev. A}\ }\textbf {\bibinfo {volume} {75}},\ \bibinfo
  {pages} {032329} (\bibinfo {year} {2007})}\BibitemShut {NoStop}%
\bibitem [{\citenamefont {Kempe}\ \emph {et~al.}(2006)\citenamefont {Kempe},
  \citenamefont {Kitaev},\ and\ \citenamefont {Regev}}]{Kempe2006}%
  \BibitemOpen
  \bibfield  {author} {\bibinfo {author} {\bibfnamefont {J.}~\bibnamefont
  {Kempe}}, \bibinfo {author} {\bibfnamefont {A.}~\bibnamefont {Kitaev}},\ and\
  \bibinfo {author} {\bibfnamefont {O.}~\bibnamefont {Regev}},\ }\href
  {https://doi.org/10.1137/S0097539704445226} {\bibfield  {journal} {\bibinfo
  {journal} {Siam J. Comput.}\ }\textbf {\bibinfo {volume} {35}},\ \bibinfo
  {pages} {1070} (\bibinfo {year} {2006})}\BibitemShut {NoStop}%
\bibitem [{\citenamefont {Cubitt}\ and\ \citenamefont
  {Montanaro}(2016)}]{Cubitt2016}%
  \BibitemOpen
  \bibfield  {author} {\bibinfo {author} {\bibfnamefont {T.}~\bibnamefont
  {Cubitt}}\ and\ \bibinfo {author} {\bibfnamefont {A.}~\bibnamefont
  {Montanaro}},\ }\href {https://doi.org/10.1137/140998287} {\bibfield
  {journal} {\bibinfo  {journal} {SIAM J. Comput.}\ }\textbf {\bibinfo {volume}
  {45}},\ \bibinfo {pages} {268} (\bibinfo {year} {2016})}\BibitemShut
  {NoStop}%
\bibitem [{\citenamefont {Piddock}\ and\ \citenamefont
  {Montanaro}(2017)}]{Piddock2017}%
  \BibitemOpen
  \bibfield  {author} {\bibinfo {author} {\bibfnamefont {S.}~\bibnamefont
  {Piddock}}\ and\ \bibinfo {author} {\bibfnamefont {A.}~\bibnamefont
  {Montanaro}},\ }\href {https://doi.org/10.26421/qic17.7-8-6} {\bibfield
  {journal} {\bibinfo  {journal} {Quantum Inf. Comput.}\ }\textbf {\bibinfo
  {volume} {17}},\ \bibinfo {pages} {636} (\bibinfo {year} {2017})}\BibitemShut
  {NoStop}%
\bibitem [{\citenamefont {Nielsen}\ and\ \citenamefont
  {Chuang}(2010)}]{Nielsen2010}%
  \BibitemOpen
  \bibfield  {author} {\bibinfo {author} {\bibfnamefont {M.~A.}\ \bibnamefont
  {Nielsen}}\ and\ \bibinfo {author} {\bibfnamefont {I.}~\bibnamefont
  {Chuang}},\ }\href@noop {} {\emph {\bibinfo {title} {{Quantum Computation and
  Quantum Information: 10th Anniversary Edition}}}}\ (\bibinfo  {publisher}
  {Cambridge University Press, Cambridge, UK},\ \bibinfo {year}
  {2010})\BibitemShut {NoStop}%
\bibitem [{\citenamefont {Walls}\ and\ \citenamefont
  {Milburn}(2008)}]{Walls2008}%
  \BibitemOpen
  \bibfield  {author} {\bibinfo {author} {\bibfnamefont {D.~F.}\ \bibnamefont
  {Walls}}\ and\ \bibinfo {author} {\bibfnamefont {G.~J.}\ \bibnamefont
  {Milburn}},\ }\href@noop {} {\emph {\bibinfo {title} {{Quantum optics}}}}\
  (\bibinfo  {publisher} {Springer Berlin Heidelberg},\ \bibinfo {year}
  {2008})\BibitemShut {NoStop}%
\bibitem [{Note1()}]{note1}%
  \BibitemOpen
  \bibinfo {note} {Detailed derivations will be provided in the Supplementary
  Information (in preparation)\label {note1}}\BibitemShut {NoStop}%
\bibitem [{\citenamefont {{Smithey, D. T. and Beck, M. and Raymer, M. G. and
  Faridani}}(1993)}]{Smithey1993}%
  \BibitemOpen
  \bibfield  {author} {\bibinfo {author} {\bibfnamefont {A.}~\bibnamefont
  {{Smithey, D. T. and Beck, M. and Raymer, M. G. and Faridani}}},\ }\href
  {https://doi.org/10.1103/PhysRevLett.70.1244} {\bibfield  {journal} {\bibinfo
   {journal} {Phys. Rev. Lett.}\ }\textbf {\bibinfo {volume} {70}},\ \bibinfo
  {pages} {1244} (\bibinfo {year} {1993})}\BibitemShut {NoStop}%
\bibitem [{\citenamefont {Lutterbach}\ and\ \citenamefont
  {Davidovich}(1997)}]{Lutterbach1997}%
  \BibitemOpen
  \bibfield  {author} {\bibinfo {author} {\bibfnamefont {L.~G.}\ \bibnamefont
  {Lutterbach}}\ and\ \bibinfo {author} {\bibfnamefont {L.}~\bibnamefont
  {Davidovich}},\ }\href {https://doi.org/10.1103/PhysRevLett.78.2547}
  {\bibfield  {journal} {\bibinfo  {journal} {Phys. Rev. Lett.}\ }\textbf
  {\bibinfo {volume} {78}},\ \bibinfo {pages} {2547} (\bibinfo {year}
  {1997})}\BibitemShut {NoStop}%
\bibitem [{\citenamefont {Bertet}\ \emph {et~al.}(2002)\citenamefont {Bertet},
  \citenamefont {Auffeves}, \citenamefont {Maioli}, \citenamefont {Osnaghi},
  \citenamefont {Meunier}, \citenamefont {Brune}, \citenamefont {Raimond},\
  and\ \citenamefont {Haroche}}]{Bertet2002}%
  \BibitemOpen
  \bibfield  {author} {\bibinfo {author} {\bibfnamefont {P.}~\bibnamefont
  {Bertet}}, \bibinfo {author} {\bibfnamefont {A.}~\bibnamefont {Auffeves}},
  \bibinfo {author} {\bibfnamefont {P.}~\bibnamefont {Maioli}}, \bibinfo
  {author} {\bibfnamefont {S.}~\bibnamefont {Osnaghi}}, \bibinfo {author}
  {\bibfnamefont {T.}~\bibnamefont {Meunier}}, \bibinfo {author} {\bibfnamefont
  {M.}~\bibnamefont {Brune}}, \bibinfo {author} {\bibfnamefont {J.~M.}\
  \bibnamefont {Raimond}},\ and\ \bibinfo {author} {\bibfnamefont
  {S.}~\bibnamefont {Haroche}},\ }\href
  {https://doi.org/10.1103/PhysRevLett.89.200402} {\bibfield  {journal}
  {\bibinfo  {journal} {Phys. Rev. Lett.}\ }\textbf {\bibinfo {volume} {89}},\
  \bibinfo {pages} {20} (\bibinfo {year} {2002})}\BibitemShut {NoStop}%
\bibitem [{\citenamefont {Krisnanda}\ \emph {et~al.}(2025)\citenamefont
  {Krisnanda}, \citenamefont {Fontaine}, \citenamefont {Copetudo},
  \citenamefont {Song}, \citenamefont {Lee}, \citenamefont {Huang},
  \citenamefont {Valadares}, \citenamefont {Liew},\ and\ \citenamefont
  {Gao}}]{Krisnanda2025}%
  \BibitemOpen
  \bibfield  {author} {\bibinfo {author} {\bibfnamefont {T.}~\bibnamefont
  {Krisnanda}}, \bibinfo {author} {\bibfnamefont {C.~Y.}\ \bibnamefont
  {Fontaine}}, \bibinfo {author} {\bibfnamefont {A.}~\bibnamefont {Copetudo}},
  \bibinfo {author} {\bibfnamefont {P.}~\bibnamefont {Song}}, \bibinfo {author}
  {\bibfnamefont {K.~X.}\ \bibnamefont {Lee}}, \bibinfo {author} {\bibfnamefont
  {N.~N.}\ \bibnamefont {Huang}}, \bibinfo {author} {\bibfnamefont
  {F.}~\bibnamefont {Valadares}}, \bibinfo {author} {\bibfnamefont {T.~C.}\
  \bibnamefont {Liew}},\ and\ \bibinfo {author} {\bibfnamefont {Y.~Y.}\
  \bibnamefont {Gao}},\ }\href {https://doi.org/10.1103/PRXQuantum.6.010303}
  {\bibfield  {journal} {\bibinfo  {journal} {PRX Quantum}\ }\textbf {\bibinfo
  {volume} {6}},\ \bibinfo {pages} {010303} (\bibinfo {year}
  {2025})}\BibitemShut {NoStop}%
\bibitem [{\citenamefont {Cahill}\ and\ \citenamefont
  {Glauber}(1969)}]{Cahill1969}%
  \BibitemOpen
  \bibfield  {author} {\bibinfo {author} {\bibfnamefont {K.~E.}\ \bibnamefont
  {Cahill}}\ and\ \bibinfo {author} {\bibfnamefont {R.~J.}\ \bibnamefont
  {Glauber}},\ }\href {https://doi.org/10.1103/PhysRev.177.1882} {\bibfield
  {journal} {\bibinfo  {journal} {Phys. Rev.}\ }\textbf {\bibinfo {volume}
  {177}},\ \bibinfo {pages} {1882} (\bibinfo {year} {1969})}\BibitemShut
  {NoStop}%
\bibitem [{\citenamefont {Haroche}\ and\ \citenamefont
  {Kleppner}(1989)}]{Haroche1989}%
  \BibitemOpen
  \bibfield  {author} {\bibinfo {author} {\bibfnamefont {S.}~\bibnamefont
  {Haroche}}\ and\ \bibinfo {author} {\bibfnamefont {D.}~\bibnamefont
  {Kleppner}},\ }\href {https://doi.org/10.1063/1.881201} {\bibfield  {journal}
  {\bibinfo  {journal} {Phys. Today}\ }\textbf {\bibinfo {volume} {42}},\
  \bibinfo {pages} {24} (\bibinfo {year} {1989})}\BibitemShut {NoStop}%
\bibitem [{\citenamefont {Raimond}\ \emph {et~al.}(2001)\citenamefont
  {Raimond}, \citenamefont {Brune},\ and\ \citenamefont
  {Haroche}}]{Raimond2001}%
  \BibitemOpen
  \bibfield  {author} {\bibinfo {author} {\bibfnamefont {J.~M.}\ \bibnamefont
  {Raimond}}, \bibinfo {author} {\bibfnamefont {M.}~\bibnamefont {Brune}},\
  and\ \bibinfo {author} {\bibfnamefont {S.}~\bibnamefont {Haroche}},\ }\href
  {https://doi.org/10.1103/RevModPhys.73.565} {\bibfield  {journal} {\bibinfo
  {journal} {Rev. Mod. Phys.}\ }\textbf {\bibinfo {volume} {73}},\ \bibinfo
  {pages} {565} (\bibinfo {year} {2001})}\BibitemShut {NoStop}%
\bibitem [{\citenamefont {Haroche}\ \emph {et~al.}(2020)\citenamefont
  {Haroche}, \citenamefont {Brune},\ and\ \citenamefont
  {Raimond}}]{Haroche2020}%
  \BibitemOpen
  \bibfield  {author} {\bibinfo {author} {\bibfnamefont {S.}~\bibnamefont
  {Haroche}}, \bibinfo {author} {\bibfnamefont {M.}~\bibnamefont {Brune}},\
  and\ \bibinfo {author} {\bibfnamefont {J.~M.}\ \bibnamefont {Raimond}},\
  }\href {https://doi.org/10.1038/s41567-020-0812-1} {\bibfield  {journal}
  {\bibinfo  {journal} {Nat. Phys.}\ }\textbf {\bibinfo {volume} {16}},\
  \bibinfo {pages} {243} (\bibinfo {year} {2020})}\BibitemShut {NoStop}%
\bibitem [{\citenamefont {Blais}\ \emph {et~al.}(2021)\citenamefont {Blais},
  \citenamefont {Grimsmo}, \citenamefont {Girvin},\ and\ \citenamefont
  {Wallraff}}]{Blais2021}%
  \BibitemOpen
  \bibfield  {author} {\bibinfo {author} {\bibfnamefont {A.}~\bibnamefont
  {Blais}}, \bibinfo {author} {\bibfnamefont {A.~L.}\ \bibnamefont {Grimsmo}},
  \bibinfo {author} {\bibfnamefont {S.~M.}\ \bibnamefont {Girvin}},\ and\
  \bibinfo {author} {\bibfnamefont {A.}~\bibnamefont {Wallraff}},\ }\href
  {https://doi.org/10.1103/RevModPhys.93.025005} {\bibfield  {journal}
  {\bibinfo  {journal} {Rev. Mod. Phys.}\ }\textbf {\bibinfo {volume} {93}},\
  \bibinfo {pages} {025005} (\bibinfo {year} {2021})}\BibitemShut {NoStop}%
\bibitem [{\citenamefont {Ma}\ \emph {et~al.}(2021)\citenamefont {Ma},
  \citenamefont {Puri}, \citenamefont {Schoelkopf}, \citenamefont {Devoret},
  \citenamefont {Girvin},\ and\ \citenamefont {Jiang}}]{Ma2021}%
  \BibitemOpen
  \bibfield  {author} {\bibinfo {author} {\bibfnamefont {W.~L.}\ \bibnamefont
  {Ma}}, \bibinfo {author} {\bibfnamefont {S.}~\bibnamefont {Puri}}, \bibinfo
  {author} {\bibfnamefont {R.~J.}\ \bibnamefont {Schoelkopf}}, \bibinfo
  {author} {\bibfnamefont {M.~H.}\ \bibnamefont {Devoret}}, \bibinfo {author}
  {\bibfnamefont {S.~M.}\ \bibnamefont {Girvin}},\ and\ \bibinfo {author}
  {\bibfnamefont {L.}~\bibnamefont {Jiang}},\ }\href
  {https://doi.org/10.1016/j.scib.2021.05.024} {\bibfield  {journal} {\bibinfo
  {journal} {Sci. Bull.}\ }\textbf {\bibinfo {volume} {66}},\ \bibinfo {pages}
  {1789} (\bibinfo {year} {2021})}\BibitemShut {NoStop}%
\bibitem [{\citenamefont {Lieb}\ \emph {et~al.}(1961)\citenamefont {Lieb},
  \citenamefont {Schultz},\ and\ \citenamefont {Mattis}}]{Lieb1961}%
  \BibitemOpen
  \bibfield  {author} {\bibinfo {author} {\bibfnamefont {E.}~\bibnamefont
  {Lieb}}, \bibinfo {author} {\bibfnamefont {T.}~\bibnamefont {Schultz}},\ and\
  \bibinfo {author} {\bibfnamefont {D.}~\bibnamefont {Mattis}},\ }\href
  {https://doi.org/10.1016/0003-4916(61)90115-4} {\bibfield  {journal}
  {\bibinfo  {journal} {Ann. Phys. (N. Y).}\ }\textbf {\bibinfo {volume}
  {16}},\ \bibinfo {pages} {407} (\bibinfo {year} {1961})}\BibitemShut
  {NoStop}%
\bibitem [{\citenamefont {Barouch}\ \emph {et~al.}(1970)\citenamefont
  {Barouch}, \citenamefont {McCoy},\ and\ \citenamefont
  {Dresden}}]{Barouch1970}%
  \BibitemOpen
  \bibfield  {author} {\bibinfo {author} {\bibfnamefont {E.}~\bibnamefont
  {Barouch}}, \bibinfo {author} {\bibfnamefont {B.~M.}\ \bibnamefont {McCoy}},\
  and\ \bibinfo {author} {\bibfnamefont {M.}~\bibnamefont {Dresden}},\ }\href
  {https://doi.org/10.1103/PhysRevA.2.1075} {\bibfield  {journal} {\bibinfo
  {journal} {Phys. Rev. A}\ }\textbf {\bibinfo {volume} {2}},\ \bibinfo {pages}
  {1075} (\bibinfo {year} {1970})}\BibitemShut {NoStop}%
\bibitem [{\citenamefont {Kosterlitz}\ and\ \citenamefont
  {Thouless}(1973)}]{Kosterlitz1973}%
  \BibitemOpen
  \bibfield  {author} {\bibinfo {author} {\bibfnamefont {J.~M.}\ \bibnamefont
  {Kosterlitz}}\ and\ \bibinfo {author} {\bibfnamefont {D.~J.}\ \bibnamefont
  {Thouless}},\ }\href {https://doi.org/10.1088/0022-3719/6/7/010} {\bibfield
  {journal} {\bibinfo  {journal} {J. Phys. C Solid State Phys.}\ }\textbf
  {\bibinfo {volume} {6}},\ \bibinfo {pages} {1181} (\bibinfo {year}
  {1973})}\BibitemShut {NoStop}%
\bibitem [{\citenamefont {Paik}\ \emph {et~al.}(2011)\citenamefont {Paik},
  \citenamefont {Schuster}, \citenamefont {Bishop}, \citenamefont {Kirchmair},
  \citenamefont {Catelani}, \citenamefont {Sears}, \citenamefont {Johnson},
  \citenamefont {Reagor}, \citenamefont {Frunzio}, \citenamefont {Glazman},
  \citenamefont {Girvin}, \citenamefont {Devoret},\ and\ \citenamefont
  {Schoelkopf}}]{Paik2011}%
  \BibitemOpen
  \bibfield  {author} {\bibinfo {author} {\bibfnamefont {H.}~\bibnamefont
  {Paik}}, \bibinfo {author} {\bibfnamefont {D.~I.}\ \bibnamefont {Schuster}},
  \bibinfo {author} {\bibfnamefont {L.~S.}\ \bibnamefont {Bishop}}, \bibinfo
  {author} {\bibfnamefont {G.}~\bibnamefont {Kirchmair}}, \bibinfo {author}
  {\bibfnamefont {G.}~\bibnamefont {Catelani}}, \bibinfo {author}
  {\bibfnamefont {A.~P.}\ \bibnamefont {Sears}}, \bibinfo {author}
  {\bibfnamefont {B.~R.}\ \bibnamefont {Johnson}}, \bibinfo {author}
  {\bibfnamefont {M.~J.}\ \bibnamefont {Reagor}}, \bibinfo {author}
  {\bibfnamefont {L.}~\bibnamefont {Frunzio}}, \bibinfo {author} {\bibfnamefont
  {L.~I.}\ \bibnamefont {Glazman}}, \bibinfo {author} {\bibfnamefont {S.~M.}\
  \bibnamefont {Girvin}}, \bibinfo {author} {\bibfnamefont {M.~H.}\
  \bibnamefont {Devoret}},\ and\ \bibinfo {author} {\bibfnamefont {R.~J.}\
  \bibnamefont {Schoelkopf}},\ }\href
  {https://doi.org/10.1103/PhysRevLett.107.240501} {\bibfield  {journal}
  {\bibinfo  {journal} {Phys. Rev. Lett.}\ }\textbf {\bibinfo {volume} {107}},\
  \bibinfo {pages} {240501} (\bibinfo {year} {2011})}\BibitemShut {NoStop}%
\bibitem [{\citenamefont {Nguyen}\ \emph {et~al.}(2019)\citenamefont {Nguyen},
  \citenamefont {Lin}, \citenamefont {Somoroff}, \citenamefont {Mencia},
  \citenamefont {Grabon},\ and\ \citenamefont {Manucharyan}}]{Nguyen2019}%
  \BibitemOpen
  \bibfield  {author} {\bibinfo {author} {\bibfnamefont {L.~B.}\ \bibnamefont
  {Nguyen}}, \bibinfo {author} {\bibfnamefont {Y.~H.}\ \bibnamefont {Lin}},
  \bibinfo {author} {\bibfnamefont {A.}~\bibnamefont {Somoroff}}, \bibinfo
  {author} {\bibfnamefont {R.}~\bibnamefont {Mencia}}, \bibinfo {author}
  {\bibfnamefont {N.}~\bibnamefont {Grabon}},\ and\ \bibinfo {author}
  {\bibfnamefont {V.~E.}\ \bibnamefont {Manucharyan}},\ }\href
  {https://doi.org/10.1103/PHYSREVX.9.041041} {\bibfield  {journal} {\bibinfo
  {journal} {Phys. Rev. X}\ }\textbf {\bibinfo {volume} {9}},\ \bibinfo {pages}
  {041041} (\bibinfo {year} {2019})}\BibitemShut {NoStop}%
\bibitem [{\citenamefont {Krantz}\ \emph {et~al.}(2019)\citenamefont {Krantz},
  \citenamefont {Kjaergaard}, \citenamefont {Yan}, \citenamefont {Orlando},
  \citenamefont {Gustavsson},\ and\ \citenamefont {Oliver}}]{Krantz2019}%
  \BibitemOpen
  \bibfield  {author} {\bibinfo {author} {\bibfnamefont {P.}~\bibnamefont
  {Krantz}}, \bibinfo {author} {\bibfnamefont {M.}~\bibnamefont {Kjaergaard}},
  \bibinfo {author} {\bibfnamefont {F.}~\bibnamefont {Yan}}, \bibinfo {author}
  {\bibfnamefont {T.~P.}\ \bibnamefont {Orlando}}, \bibinfo {author}
  {\bibfnamefont {S.}~\bibnamefont {Gustavsson}},\ and\ \bibinfo {author}
  {\bibfnamefont {W.~D.}\ \bibnamefont {Oliver}},\ }\href
  {https://doi.org/10.1063/1.5089550} {\bibfield  {journal} {\bibinfo
  {journal} {Appl. Phys. Rev.}\ }\textbf {\bibinfo {volume} {6}},\ \bibinfo
  {pages} {021318} (\bibinfo {year} {2019})}\BibitemShut {NoStop}%
\bibitem [{Note2()}]{note2}%
  \BibitemOpen
  \bibinfo {note} {Defined as respectively\label {note2}}\BibitemShut {NoStop}%
\bibitem [{\citenamefont {Johansson}\ \emph {et~al.}(2012)\citenamefont
  {Johansson}, \citenamefont {Nation},\ and\ \citenamefont
  {Nori}}]{Johansson2012}%
  \BibitemOpen
  \bibfield  {author} {\bibinfo {author} {\bibfnamefont {J.~R.}\ \bibnamefont
  {Johansson}}, \bibinfo {author} {\bibfnamefont {P.~D.}\ \bibnamefont
  {Nation}},\ and\ \bibinfo {author} {\bibfnamefont {F.}~\bibnamefont {Nori}},\
  }\href {https://doi.org/10.1016/j.cpc.2012.02.021} {\bibfield  {journal}
  {\bibinfo  {journal} {Comput. Phys. Commun.}\ }\textbf {\bibinfo {volume}
  {183}},\ \bibinfo {pages} {1760} (\bibinfo {year} {2012})}\BibitemShut
  {NoStop}%
\bibitem [{\citenamefont {Johansson}\ \emph {et~al.}(2013)\citenamefont
  {Johansson}, \citenamefont {Nation},\ and\ \citenamefont
  {Nori}}]{Johansson2013}%
  \BibitemOpen
  \bibfield  {author} {\bibinfo {author} {\bibfnamefont {J.~R.}\ \bibnamefont
  {Johansson}}, \bibinfo {author} {\bibfnamefont {P.~D.}\ \bibnamefont
  {Nation}},\ and\ \bibinfo {author} {\bibfnamefont {F.}~\bibnamefont {Nori}},\
  }\href {https://doi.org/10.1016/j.cpc.2012.11.019} {\bibfield  {journal}
  {\bibinfo  {journal} {Comput. Phys. Commun.}\ }\textbf {\bibinfo {volume}
  {184}},\ \bibinfo {pages} {1234} (\bibinfo {year} {2013})}\BibitemShut
  {NoStop}%
\bibitem [{Note3()}]{note3}%
  \BibitemOpen
  \bibinfo {note} {Comprehensive simulation details will be provided in the
  Supplementary Information (in preparation)\label {note3}}\BibitemShut
  {NoStop}%
\bibitem [{\citenamefont {Farhi}\ \emph {et~al.}(2001)\citenamefont {Farhi},
  \citenamefont {Goldstone}, \citenamefont {Gutmann}, \citenamefont {Lapan},
  \citenamefont {Lundgren},\ and\ \citenamefont {Preda}}]{Farhi2001}%
  \BibitemOpen
  \bibfield  {author} {\bibinfo {author} {\bibfnamefont {E.}~\bibnamefont
  {Farhi}}, \bibinfo {author} {\bibfnamefont {J.}~\bibnamefont {Goldstone}},
  \bibinfo {author} {\bibfnamefont {S.}~\bibnamefont {Gutmann}}, \bibinfo
  {author} {\bibfnamefont {J.}~\bibnamefont {Lapan}}, \bibinfo {author}
  {\bibfnamefont {A.}~\bibnamefont {Lundgren}},\ and\ \bibinfo {author}
  {\bibfnamefont {D.}~\bibnamefont {Preda}},\ }\href
  {https://doi.org/10.1126/science.1057726} {\bibfield  {journal} {\bibinfo
  {journal} {Science}\ }\textbf {\bibinfo {volume} {292}},\ \bibinfo {pages}
  {472} (\bibinfo {year} {2001})}\BibitemShut {NoStop}%
\end{thebibliography}%

\end{document}